\documentstyle[12pt,psfig]{article}

\begin{document}

\baselineskip=.22in
\renewcommand{\baselinestretch}{1.2}
\renewcommand{\theequation}{\thesection.\arabic{equation}}

\begin{flushright}
hep-th/0212326
\end{flushright}

\vspace{5mm}

\begin{center}
{\Large \bf Classical Geometry of De Sitter Spacetime :\\[2mm]
An Introductory Review}\\[12mm]
Yoonbai Kim${}^{a,b,}$\footnote{yoonbai$@$skku.ac.kr}, 
Chae Young Oh${}^{a,}$\footnote{young$@$newton.skku.ac.kr}, 
and Namil Park${}^{a,}$\footnote{namil$@$newton.skku.ac.kr}
\\[5mm] \em
${}^{a}$BK21 Physics Research Division and Institute of Basic Science,
Sungkyunkwan University,\\
Suwon 440-746, Korea\\[2mm]
${}^{b}$School of Physics, Korea Institute for Advanced Study,\\
207-43, Cheongryangri-Dong, Dongdaemun-Gu, Seoul 130-012, Korea
\end{center}

\vspace{5mm}

\begin{abstract}
Classical geometry of de Sitter spacetime is reviewed in arbitrary dimensions.
Topics include coordinate systems, geodesic motions, and Penrose diagrams
with detailed calculations.
\end{abstract}

\newpage

\tableofcontents

\setcounter{equation}{0}
\section{Introduction}

Cosmological constant $\Lambda$ was originally introduced in general
relativity with the motivation to allow static homogeneous universe
to Einstein equations in the presence of matter. Once expansion of the 
universe was discovered, its role turned out to be unnecessary and
it experienced a checkered history. Recent cosmological observations
suggest some evidences for the existence of a positive cosmological
constant~\cite{Rie}~: From a variety of observational sources, its 
magnitude is $(G\hbar/c^{3})\Lambda\sim 10^{-123}$, and both matter 
$\Omega_{\rm matter}$ and vacuum $\Omega_{\Lambda}$ are of comparable
magnitude, $\Omega_{\rm matter}/\Omega_{\Lambda}\sim 0.3/0.7$ in 
spatially flat universe. Both results are paraphrased as old and new
cosmological constant problems~\cite{old,new}. 
Inclusion of quantum field theory predicts
naturally the energy density via quantum fluctuations even up to
the Planck scale, so the old problem is to understand why the 
observed present cosmological constant is so small. It is the worst 
fine-tuning problem requiring a magic cancellation with accuracy to
120 decimal places. The new problem is to understand why the vacuum
energy density is comparable to the present mass density of 
the universe. Though there are several classes of efforts to solve
the cosmological constant problem, e.g., cancellation of vacuum energy
based on symmetry principle like supersymmetry, the idea of quintessence,
and the anthropic principle, any of them is not widely accepted as the 
solution, 
yet. 
In the early universe, existing cosmological constant also provides 
an intriguing idea, inflationary cosmology, to fix long-standing 
cosmological problems in the Big-bang scenario.

In addition to the above, detailed properties of the de Sitter
geometry in arbitrary dimensions  
raise a debate like selection
of true vacuum compatible with its scale symmetry in relation
with trans-Planckian cosmology~\cite{tran,covac} or becomes a cornerstone
of a theoretical idea like dS/CFT for a challenging problem of quantum 
gravity in de Sitter spacetime~\cite{dscft}.

In the context of general relativity, classical study 
on the cosmological constant is to understand geometry of de Sitter 
spacetime (dS${}_d$). 
Therefore, in this review, we study de Sitter geometry and the motion
of classical test particle in the background of the de Sitter spacetime
in detail. Topics include useful coordinate systems, geodesic
motions, and Penrose diagrams. One of our 
purposes is to combine knowledges of classical 
de Sitter geometry, scattered with other subjects in the textbooks or 
reviews, in one note~\cite{BD,KT,text,revi}. We tried to make our review
self-contained and a technical note by containing various formulas
in the appendix.

The review is organized as follows. In section 2, we introduce our setup
with notations and signature. In section 3, four coordinates (global,
conformal, planar, and static) and coordinate transformations among those
are explained with the reason why such four of them are frequently used. 
Killing symmetries are also obtained in each coordinate system.
All possible geodesic motions of classical test particle 
are given in section 4. Identification of de Sitter horizon for a static
observer and cosmological evolution to a comoving observer
are also included. In section 5, causal structure of the global
de Sitter spacetime is dealt through Penrose diagrams, and its quantum
theoretical implication is also discussed. Section 6 is
devoted to brief discussion. Various quantities are displayed 
in Appendix for convenience. 

\setcounter{equation}{0}
\section{Setup}
In this section, we introduce basic setup for studying classical geometry 
of de Sitter spacetime in arbitrary dimensions.
Two methods are employed : One is solving directly Einstein equations
for each metric ansatz, and the other is reading off the specific form of
coordinate transformation between two metrics.

We begin with Einstein-Hilbert action coupled to matters~:
\begin{equation}\label{act}
S=\frac{1}{16\pi G}\int d^d x\, \sqrt{-g}\, (R-2\Lambda)+S_{\rm m},
\end{equation}
where $S_{\rm m}$ stands for the matter action of our interest, which 
vanishes for limit of the pure gravity, and $\Lambda$ is a cosmological
constant, which sets to be positive for dS${}_d$.
Einstein equations are read from the action (\ref{act})
\begin{equation}\label{eeq}
G_{\mu\nu}+ \Lambda g_{\mu\nu}= 8 \pi G T_{\mu\nu},
\end{equation}
where energy-momentum tensor $T_{\mu\nu}$ is defined by
\begin{eqnarray}
T_{\mu\nu}\equiv  -\frac{2}{\sqrt{-g}}\frac{\delta {S}_{\rm
m}}{\delta g^{\mu\nu}}.
\label{tmn}
\end{eqnarray}
Throughout the paper small Greek indices $\mu,\nu,\rho,\sigma,\tau,\kappa...$
 run from $0$ to $d-1$, and our spacetime signature is $(-,+,+,\dots,+)$.

To be specific let us consider a real scalar field $\phi(x)$ of which
action is
\begin{equation}
S_{\rm m}=\int d^dx\sqrt{-g}\left[-\frac{1}{2}g^{\mu\nu}\partial_\mu\phi
\partial_\nu\phi-V(\phi)\right].
\end{equation}
In order not to have an additional contribution to the cosmological constant at 
classical level, we set minimum value of the scalar potential to vanish,
${\rm min}(V(\phi))=0$. 
Then its energy-momentum tensor becomes 
\begin{equation}
T_{\mu\nu}=\partial_\mu\phi \partial_\nu \phi-\frac{1}{2}g_{\mu\nu}
\partial_\rho\phi \partial^\rho \phi-g_{\mu\nu}V(\phi),
\end{equation}
where energy density $T_{00}$ is positive semi-definite in the limit of 
flat spacetime.

For the pure dS${}_d$, the energy-momentum tensor of the matter, $T_{\mu\nu}$, 
vanishes so that one can regard these spacetimes as solutions of the Einstein
equations of Eq.~(\ref{eeq})
\begin{equation}\label{2}
G_{\mu\nu}=- \Lambda g_{\mu\nu}
\end{equation}
for an empty spacetime with a positive constant vacuum energy $(\Lambda >0)$ :
\begin{equation}\label{sour}
T_{\mu\nu}^{\rm vacuum}\equiv\frac{\Lambda}{8 \pi G}g_{\mu\nu}.
\end{equation}
Therefore, the only nontrivial component of the Einstein equations (\ref{2}) is
\begin{equation}\label{4}
R = \frac {2d}{d-2} \Lambda > 0.
\end{equation}
It means that the de Sitter spacetime is maximally symmetric, of which
local structure is characterized by a positive constant curvature scalar 
alone such as
\begin{equation}\label{9}
R_{\mu\nu\rho\sigma} = \frac {1}{d(d-1)} ( g_{\mu\rho} g_{\nu\sigma} -
g_{\mu\sigma} g_{\nu\rho} ) R .
\end{equation}
Since the scalar curvature (\ref{4}) is constant everywhere, 
the dS${}_{d}$ is free from physical singularity and it is 
confirmed by a constant Kretschmann scalar :
\begin{equation}
R_{\mu\nu\rho\sigma}R^{\mu\nu\rho\sigma}=\frac{2}{d(d-1)}R^{2}
=\frac{8d^2}{d(d-1)(d-2)^2}\Lambda^2.
\end{equation}

We follow usual definition of tensors as follows.
\begin{center}
\renewcommand{\arraystretch}{1.7}
\begin{tabular}{|c |c |}\hline
 quantity & definition \\ \hline
Jacobian factor & $ g \equiv \det(g_{\mu\nu}) $ \\
connection           &   $\Gamma ^\mu _{\nu\rho} \equiv
\frac{1}{2}g^{\mu\sigma}(\partial_\nu g_{\sigma \rho}+\partial_\rho
g_{\sigma \nu} -\partial_\sigma g_{\nu \rho}) $          \\
covariant derivative of a contravariant vector & $ \nabla_\mu A^\nu
\equiv \partial_\mu A^\nu + \Gamma^\nu_{\mu\rho}A^\rho $      \\

Riemann curvature tensor & $
R^{\mu}_{\;\;\nu\rho\sigma} \equiv \partial_\rho \Gamma^{\mu}_{\sigma\nu}
-\partial_{\sigma} \Gamma^{\mu}_{\rho\nu}+
\Gamma^{\mu}_{\rho\tau}\Gamma^{\tau}_{\sigma\nu}
-\Gamma^{\mu}_{\sigma\tau} \Gamma^{\tau}_{\rho\nu} $\\
Ricci tensor  &  $ R_{\mu\nu} \equiv R^{\rho}_{ \; \mu\rho\nu}$ \\
curvature scalar  & $ R \equiv g^{\mu\nu}R_{\mu\nu}$ \\
Einstein tensor & $\displaystyle{G_{\mu\nu}\equiv R_{\mu\nu}
-\frac{g_{\mu\nu}}{2}R}$ \\
\hline
\end{tabular}
\end{center}

\setcounter{equation}{0}
\section{Useful Coordinates}
When we deal with specific physics in general relativity, we are firstly
concerned with appropriate coordinate system.
In this section, we introduce four useful coordinates and build
coordinate transformations among those for tensor calculus~: They are 
global (closed), conformal, planar (inflationary), and static coordinates. 
For the detailed calculation of various quantities in systems of 
coordinates, refer to Appendix.

Similar to physical issues in flat spacetime, the symmetries of a Riemannian
manifold is realized through metric invariance under the symmetry 
transformations. Specifically, let us choose a coordinate system  
as $(x^{\mu},g_{\mu\nu})$ and take into account a transformation 
$x^{\mu}\rightarrow x'^{\mu}$. If the metric $g_{\mu\nu}$ remains to be 
form-invariant under the transformation
\begin{equation}\label{fin}
g_{\mu\nu}'(x)=g_{\mu\nu}(x)
\end{equation}
for all coordinates $x^{\mu}$, such transformation is called an isometry.
Since any finite continuous transformation with non-zero Jacobian 
can be constructed by an infinite sum of infinitesimal transformations,
study under infinitesimal transformation is sufficient for continuous 
isometry. If we consider an infinitesimal coordinate transformation
\begin{equation}
x^{\mu}\rightarrow x'^{\mu}=x^{\mu}+\varepsilon X^{\mu}(x),
\end{equation}
where $\varepsilon$ is small parameter and $X^{\mu}$ a vector field, 
expansion of the form-invariance (\ref{fin}) leads to Killing's equations
\begin{eqnarray}\label{Keq}
L_{X}g_{\mu\nu}\equiv X^{\rho}\partial_{\rho}g_{\mu\nu}
+g_{\mu\rho}\partial_{\nu}X^{\rho}
+g_{\rho\nu}\partial_{\mu}X^{\rho}=0,
\end{eqnarray} 
where $L_{X}$ denotes a Lie derivative.
Then, any $X^{\mu}$ given by a solution of the Killing's equations (\ref{Keq})
is called a Killing vector field. In every subsection we will discuss
Killing symmetries of each coordinate system.

\subsection{Global (closed) coordinates
$(\tau , \theta_{i}),~i=1,2,\cdots ,d-1$}
An intriguing observation about the dS${}_{d}$ is embedding of the
dS${}_{d}$
into flat $(d+1)$-dimensional spacetime, which was made by E.
Schr\"{o}dinger (1956).
In $(d+1)$-dimensional Minkowski spacetime, the Einstein equation is
trivially satisfied 
\begin{eqnarray}\label{10}
0 &=& ^{d+1}R \equiv g^{AB} R_{AB} \nonumber\\
&=& R + R_{dd} ,
\end{eqnarray}
where capital indices, $A,B,...$, represent $(d+1)$-dimensional
Minkowski indices run from $0$ to $d$. If we set 
$R_{dd}=-\frac{2d}{d-2} \Lambda$
which implies a positive constant curvature of the embedded space, then
the $d$-dimensional Einstein equation (\ref{4}) of dS${}_{d}$ is 
reproduced. 

Topology of such embedded $d$-dimensional spacetime of the positive 
constant curvature is visualized by urging algebraic constraint 
of a hyperboloid
\begin{equation}\label{12}
\eta _{AB} X^A X^B = l^2 ,
\end{equation}
where $\eta _{AB} = {\rm diag}( -1, 1, 1, \cdot \cdot \cdot ,1)$.
So the globally defined metric of flat $(d+1)$-dimensional spacetime is
\begin{equation}\label{11}
ds^2=\eta_{AB}dX^A dX^B
\end{equation}
constrained by the equation of a hyperboloid (\ref{12}).
Let us clarify the relation between the cosmological constant $\Lambda$ and the
length $l$ by inserting the constraint (\ref{12}) into the Einstein equation
(\ref{4}). To be specific, we eliminate the last spatial coordinate $X^{d}$
from the metric (\ref{11}) and the constraint (\ref{12}) :
\begin{equation}\label{xd1}
dX^d =\mp \frac{\eta_{\mu\nu}X^{\mu} dX^{\nu}}{\sqrt{l^2-\eta_{\alpha
 \beta} X^{\alpha} X^{\beta}}}.
\end{equation}
Inserting Eq.~(\ref{xd1}) into the metric (\ref{11}), we have the induced
metric $g_{\mu\nu}$ of a curved spacetime due to an embedding :
\begin{equation}\label{gbmn}
g_{\mu\nu}= \eta _{\mu\nu} +
\frac{X_\mu X_\nu }{l^2-\eta _{\alpha\beta}X^\alpha X^\beta},
\end{equation}
and its inverse $g^{\mu\nu}$ is
\begin{equation}\label{gtmn}
g^{\mu\nu}= \eta ^{\mu \nu} - \frac{1}{l^2} X^\mu X^\nu.
\end{equation}
The induced connection 
$\Gamma ^\mu _{\nu\rho}$ becomes
\begin{eqnarray}\label{csym}
\Gamma^{\mu}_{\nu\rho}= \frac{1}{l^2}
\left(\eta_{\nu \rho} X^\mu
+ \frac{X_\nu X_\rho X^\mu}{l^2-\eta _{\alpha\beta}X^\alpha X^\beta}
\right).
\end{eqnarray}
Subsequently, we compute induced curvatures : The Riemann curvature tensor
$R^{\mu}_{\;\nu\rho\sigma}$ is
\begin{eqnarray}\label{rie}
R^{\mu}_{\;\nu\rho\sigma}=
\frac{1}{l^2} \left[ \eta _{\nu \sigma} \delta {^\mu}{_\rho}
- \eta _{\nu \rho} \delta {^\mu}{_\sigma}
+ \frac{\eta_{\nu \rho} X_\sigma X^\mu - \eta_{\nu \sigma}X_\rho X^\mu
+ \delta {^\mu}{_\rho} X_\nu X_\sigma
- \delta {^\mu}{_\sigma} X_\nu X_\rho }
{l^2-\eta _{\alpha\beta}X^\alpha X^\beta}
\right] \nonumber\\
+ \frac{1}{l^4} \left[ \eta_{\nu \sigma} X_\rho X^\mu
- \eta_{\nu \rho} X_\sigma X^\mu
+ \frac{ \left(\eta_{\nu \sigma} X_\rho X^\mu
- \eta_{\nu \rho} X_\sigma X^\mu \right) X_\gamma X^\gamma}
{l^2-\eta _{\alpha\beta}X^\alpha X^\beta}
\right],~~~~~~~~~
\end{eqnarray}
and the Ricci tensor $R_{\mu\nu}$ is 
\begin{equation}\label{rict}
R_{\mu\nu} = \frac{d-1}{l^2} \left( \eta_{\mu \nu}
+ \frac{X_\mu X_\nu}{l^2-\eta_{\alpha \beta}X^\alpha X^\beta} \right)\;.
\end{equation}
Since the induced curvature scalar 
$R$ does not vanish
\begin{eqnarray}\label{rics}
R= \frac{d(d-1)}{l^2},
\end{eqnarray}
the Einstein equation, (\ref{10}) or (\ref{12}), will determine the relation
between the length $l$ and the cosmological constant $\Lambda$. 
Insertion of the constraint (\ref{12}) into the Einstein equation
(\ref{4})
gives us a relation between the cosmological constant $\Lambda$ and the
length $l$ such as
\begin{eqnarray}\label{len}
\Lambda=\frac{(d-1)(d-2)}{2l^{2}}.
\end{eqnarray}

\begin{figure}[ht]
\centerline{\psfig{figure=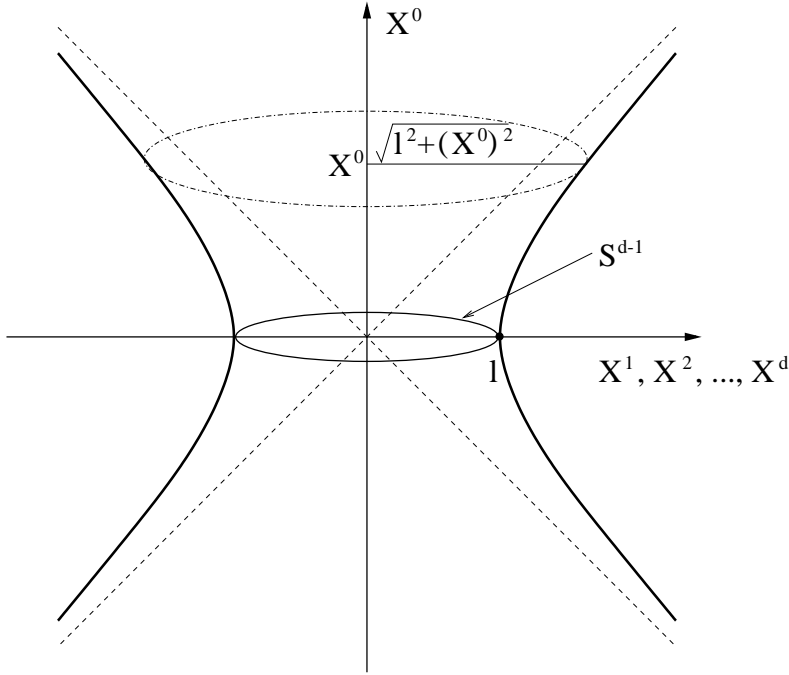,height=90mm}}
\caption{$d$-dimensional hyperboloid illustrating de Sitter spacetime
embedded in $d+1$-dimensions.}
\label{dsrfig1}
\end{figure}

Implemented by Weyl symmetry, the unique scale of the theory of our
interest, $l$,
can always be set to be one by a Weyl rescaling as
\begin{equation}\label{wsc}
X^A \longrightarrow X ^{'{A}} = e^{-\omega} X^A =\frac{X^A}{l} .
\end{equation}
Therefore, the dS${}_{d}$ is defined as a $d$-dimensional hypersurface
(a hyperboloid) embedded in flat $(d+1)$-dimensional spacetime, and
the overall topology is cylindrical, being $R\times S^{d-1}$ (See
Fig.~\ref{dsrfig1}).

A well-known coordinate system to cover entire $d$-dimensional 
hyperboloid is obtained from the following observation : 
As shown in Fig.~\ref{dsrfig1}, the relation between $X^{0}$ and 
the spatial length of $(X^{1},X^{2},...,X^{d})$ is hyperbolic 
and spatial sections of a constant $X^0$ defines a $(d-1)$-dimensional 
sphere of radius $\sqrt{l^{2}+(X^{0})^{2}}$. 
Therefore, a convenient choice satisfying Eq.~(\ref{12}) is
\begin{eqnarray}\label{gc}
X^0 = l \sinh \frac{\tau}{l}~~{\rm and}~~
X^\alpha = l \omega^\alpha\cosh \frac{\tau}{l},~~(\alpha=1,...,d),
\end{eqnarray}
where $-\infty < \tau < \infty$ and $\omega^{\alpha}$'s$~(\alpha,\beta,...)$
 for the spatial sections of constant $\tau$ satisfy
\begin{equation}
\sum_{\alpha=1}^d(\omega ^\alpha)^2 = 1.
\end{equation}
Subsequently, we have $d-1$ angle variables $\theta_{i},~i=1,2,\cdots ,d-1$,
such as
\begin{eqnarray}\label{omega}
\omega ^1 &=& \cos \theta_1 ,\hspace{55mm}
0 \leq \theta {_1} < \pi, \nonumber\\
\omega ^2 &=& \sin \theta_1 \cos \theta_2 ,\hspace{45mm}
0 \leq \theta {_2} < \pi, \nonumber\\
&\vdots& \nonumber\\
\omega ^{d-2} &=& \sin \theta_1 \cos \theta_2
\cdots \sin \theta_{d-3} \cos \theta_{d-3}, ~\hspace{10mm}
0 \leq \theta_{d-3} < \pi, \nonumber\\
\omega ^{d-1} &=& \sin \theta_1 \cos \theta_2
\cdot \cdot \cdot \cdots\sin \theta_{d-2} \cos\theta_{d-1},
\hspace{5mm} 0 \leq \theta_{d-2} < \pi, \nonumber\\
\omega ^d &=& \sin \theta_1 \cos\theta_2 ,
\cdots\cdots \sin \theta_{d-2} \sin\theta_{d-1},
\hspace{5mm} 0 \leq \theta_{d-1} < 2\pi.
\end{eqnarray}
Inserting Eq.~(\ref{gc}) and Eq.~(\ref{omega}) into the metric (\ref{11}),
 we have a resultant metric rewritten in terms of the above coordinates
$(\tau , \theta_{i})$ :
\begin{equation}\label{gcs}
ds^2 =-d\tau^{2}+ l^2 \cosh ^2 \left(\frac{\tau}{l}\right)\, d\Omega^{2}_{d-1},
\end{equation}
where $d\Omega^{2}_{d-1}$ is $(d-1)$-dimensional solid angle 
\begin{eqnarray}\label{dome}
d\Omega^{2}_{d-1} &=& d\theta _1 ^2
+ \sin ^2 \theta _1 d\theta _2 ^2 + \cdot\cdot\cdot
+ \sin ^2 \theta _1 \cdot\cdot\cdot
\sin ^2 \theta_{d-2} d\theta^{2}_{d-1} \nonumber\\
&=& \sum_{j=1}^{d-1} \left(\prod _{i=1}^{j-1} \sin ^2 \theta _{i}\right)
d\theta _j ^2.
\end{eqnarray}
Note that singularities in the metric~(\ref{dome}) at $\theta_i=0$
 and $\theta_i=\pi\;(i=1,2,\cdots,d-2)$ are simply trivial coordinate 
artifacts that occur with polar coordinates.
In these coordinates with a fixed time $\tau$, the spatial hypersurface
corresponds to a $(d-1)$-dimensional sphere of radius $l\cosh (\tau/l)$.
Thus, its radius is infinitely large at $\tau=-\infty$, decreases to the 
minimum radius $l$ at $\tau=0$, and then increases to infinite size as 
$\tau \rightarrow\infty$. We again emphasize that the spatial section is
compact (finite) except for the farthest past and future. 
 In this coordinate system, 
$\partial/\partial\theta_{d-1}$ is the only Killing vector because the 
metric (\ref{gcs}) is isometric under the rotation of a coordinate 
$\theta_{d-1}$.
On the other hand, $\partial/\partial \tau$ is not a Killing vector and
 missing of this Killing symmetry breaks the conservation of energy so 
that Hamiltonian is not defined properly and the quantization procedure 
does not proceed smoothly. However, $S$-matrix of the quantum
field theory, if it exists, is known to be unitary because the coordinates
cover globally the whole spacetime.

An opposite direction is to assume an appropriate form of metric with  
an unknown function $f(\tau/l)$, 
\begin{equation}\label{gcm}
ds^{2}=-d\tau^{2}+l^{2}f(\tau/l)^{2}d\Omega^{2}_{d-1},
\end{equation}
and then to try to solve the Einstein equations.
Curvature scalar of the metric (\ref{gcm}) is computed as
\begin{equation}\label{csgo}
R=(d-1)\frac{(d-2)(1+\dot{f}^2)+2f\ddot{f}}{l^2f^2},
\end{equation}
where the overdot denotes derivative of the rescaled time variable
$\dot{f}\equiv df/d(\tau/l)$ in this subsection. 
With the aid of Eq.~(\ref{len}), the Einstein equation (\ref{4}) becomes
\begin{equation}\label{gneq}
2(f\ddot{f}-\dot{f}^2-1)=d(-\dot{f}^2+f^2-1).
\end{equation}
A particular solution of Eq.~(\ref{gneq}), irrespective of $d$, 
can be obtained by solving two equations
\begin{equation}\label{triv}
f\ddot{f}-\dot{f}^2-1=0~~\mbox{and}~~-\dot{f}^2+f^2-1=0.
\end{equation}
The solution of Eq.~(\ref{triv}) is $\pm \cosh[(\tau-\tau_0)/l]$  
except unwanted trivial solution ($f=0$). Since we can choose $\tau_0=0$ 
without loss of generality, via a time translation, 
$\tau\rightarrow\tau-\tau_0$, we have  
\begin{equation}\label{solg}
f(\tau/l)=\pm \cosh\left(\frac{\tau}{l}\right)
\end{equation}
which is equivalent to the metric (\ref{gcs}) found by a coordinate 
transformation.
The general solution of Eq.~(\ref{gneq}) should satisfy the following 
equations
\begin{equation}
f\ddot{f}-\dot{f}^2-1=d{\cal F} ~~\mbox{and}~~
-\dot{f}^2+f^2-1=2{\cal F},
\end{equation}
where ${\cal F}$ is an arbitrary function of $\tau/l$.
According to the search by use of Mathematica and a handbook~\cite{PZ},
 it seems that no other solution than Eq.~(\ref{solg}) is known, yet.

\subsection{Conformal coordinates $(T,\theta_{i})$}
A noteworthy property of the dS${}_d$ is provided by computation of
the Weyl (or conformal) tensor
\begin{eqnarray}\label{5}
C_{\mu\nu\rho\sigma} &=& R_{\mu\nu\rho\sigma} + \frac {1}{d-2}
(g_{\mu\sigma} R_{\rho\nu}
+ g_{\nu\rho} R_{\sigma\mu} - g_{\mu\rho} R_{\rho\nu} - g_{\nu\sigma}
R_{\rho\nu}) \nonumber\\
&&+\frac {1}{(d-1)(d-2)} ( g_{\mu\rho} g_{\rho\nu} - g_{\mu\sigma}
g_{\sigma\nu})R ,
\end{eqnarray}
for $d \geq 4$, or the Cotton tensor
\begin{equation}\label{6}
C^{\mu\nu} = \frac {\epsilon ^{\rho\sigma\mu}} {\sqrt {-g}}
\nabla_{\sigma}
\left( R {^\nu}{_\rho} - \frac {\delta^\nu_{\; \rho}}{4} R \right)
\end{equation}
for $d=3$.\footnote{The Cotten tensor (\ref{6}) is derived from
the gravitational Chern-Simon action :
\begin{eqnarray}\label{cs}
S_{\rm CS}=\frac{1}{2}\int d^3 x\, \epsilon^{\mu\nu\rho}
\left(R{^\rho}_{\sigma\mu\nu}
\Gamma^{\sigma} _{\rho\tau} - \frac{2}{3} \Gamma^{\rho} _{\sigma\mu}
\Gamma ^{\sigma} _{\kappa\nu} \Gamma ^{\kappa}_{\rho\tau} \right) .
\end{eqnarray}
Since this gravitational Chern-Simon action is composed of cubic
derivative
terms so that it is dimensionless without any dimensionful constant.
It means that 3-dimensional conformal gravity is described by 
this Chern-Simon term.}
Substituting the Einstein equation (\ref{4}) and Eq.~(\ref{9}) into
the conformal tensors in Eqs.~(\ref{5}) and (\ref{6}), we easily notice
that they vanish so that dS${}_d$ is proven to be
conformally flat :
\begin{eqnarray}\label{8}
C_{\mu\nu\rho\sigma} &=& \left[\frac{1}{d(d-1)} - \frac{2}{d(d-2)}
+\frac{1}{(d-1)(d-2)}\right] ( g_{\mu\rho} g_{\rho\nu} - g_{\mu\sigma}
g_{\rho\nu} )R
=0,\\
C^{\mu\nu} &=&
-\frac{1}{4}\frac{\epsilon^{\mu\nu\rho}}{\sqrt {-g}} \nabla_{\rho}
(R - 4 \Lambda ) \stackrel{R\sim\Lambda}{=}0.
\end{eqnarray}
Therefore, the condition (\ref{9}) that the Riemann curvature tensor
is determined by the scalar curvature alone is equivalent to the 
condition of vanishing conformal tensor in this system.
Consequently the unique scale of the 
dS${}_{d}$, $l=\sqrt{(d-1)(d-2)/2\Lambda}$,
can be scaled away by a Weyl (scale) transformation.

The above conformal property of the dS${}_{d}$ suggests that
conformal coordinate system is a good coordinate system. It is written in
terms of conformal time $T$ as
\begin{equation}\label{cmet}
ds^{2}=F(T/l)^{2}(-dT^{2}+l^{2}d\Omega^2_{d-1}).
\end{equation}
If we compare the metric of the global coordinates (\ref{gcs}) with
that of the conformal coordinates (\ref{cmet}), then coordinate transformation
between two is summarized in a first-order differential equation of $F(T/l)$
\begin{equation}\label{foe}
\frac{d\ln F}{dT}=\pm\sqrt{F^{2}-1}\; ,
\end{equation}
where $F(T/l)=\cosh (\tau/l)\ge 1$ provides a boundary condition such as
$F(0)=1$. The unique solution of Eq.~(\ref{foe}) is
\begin{equation}\label{usol}
F(T/l)=\sec\left(\frac{T}{l}\right).
\end{equation}
Substituting the result (\ref{usol}) into the metric (\ref{cmet}), we have
\begin{equation}\label{cc}
ds^{2}=\frac {1}{\cos^2 \frac{T}{l}} (-dT^{2}+l^2 d\Omega_{d-1} ^2)
~\mbox{with}~ - \frac{\pi}{2} < \frac{T}{l} < \frac{\pi}{2}.
\end{equation}

Since the metric (\ref{cc}) is isometric under the rotation of 
 $\theta_{d-1}$, $\partial/ \partial\theta_{d-1}$ is a Killing vector but 
there is no other Killing vector. Thus, the only 
symmetry is axial symmetry.
Note that there is one-to-one correspondence between the global coordinates
 (\ref{gcs}) and the conformal coordinates (\ref{cc}), 
which means that the conformal coordinate
system describes also entire de Sitter spacetime. In addition, any null 
geodesic
with respect to the conformal metric (\ref{cc}) is also null in the
conformally-transformed metric :
\begin{equation}\label{cct}
d\tilde{s}^{2}\equiv \cos^2 \left(\frac{T}{l}\right) ds^{2}
=-dT^{2}+l^2 d\Omega_{d-1} ^2 .
\end{equation}
Therefore, Penrose diagram for the dS${}_{d}$ can easily be identified
from this metric (\ref{cct}), which contains the whole information about the
causal structure of the dS${}_{d}$ but distances are highly
distorted. However, we will discuss the Penrose diagram for the dS${}_{d}$
in section \ref{pendi} after the Kruskal coordinate system is introduced. 
As mentioned previously, 
topology of the de Sitter spacetime is cylindrical $(R\times S^{d-1})$
so the process to
make the Penrose diagram is to change the hyperboloid into a $d$-dimensional 
cylinder of a finite height, being $I\times S^{d-1}$ $(I=[0,\pi])$.
Again note that the conformal coordinate system (\ref{cc}) does not include 
the timelike Killing symmetry so that the Hamiltonian as a conserved 
quantity cannot be chosen and the quantum theory on these coordinates 
is also sick. Existence of conformal symmetry also affects much on the
 choice of field theoretic vacuum in the de Sitter spacetime.

From now on let us obtain the conformal metric (\ref{usol}) by solving 
the Einstein equations (\ref{2}) under Eq.~(\ref{cmet}). 
If we compute the curvature scalar, we have
\begin{equation}\label{csco}
R=(d-1)\frac{(d-2)F^2+(d-4)\dot{F}^2+2F\ddot{F}}{l^2F^4},
\end{equation}
Here overdot denotes  $\dot{F} \equiv dF/d(T/l)$, and the same overdot in 
every subsection will also be used as derivative of the rescaled time
variable in each corresponding subsection. 
Then the Einstein equation (\ref{4}) becomes
\begin{equation}\label{einc}
2(F\ddot{F}-F^2-2\dot{F}^2)=d(F^4-\dot{F}^2-F^2).
\end{equation}
A particular solution of Eq.~(\ref{einc}), irrespective of $d$,
should satisfy the following equations 
\begin{equation}\label{cneq}
F\ddot{F}-2\dot{F}^2-F^2=0 ~~\mbox{and}~~\dot{F}^2+F^2-F^4=0.
\end{equation}
The unique solution of Eq.~(\ref{cneq}) with $F(0)=1$ is proven to be the
same as Eq.~(\ref{usol}).  
The general solution should satisfy the following equations
\begin{equation}
F\ddot{F}-F^2-2\dot{F}^2=d{\cal G} ~~\mbox{and}~~F^4-\dot{F}^2-F^2
= 2{\cal G},
\end{equation}
where ${\cal G}$ is an arbitrary function of $T/l$.
According to the search by use of Mathematica and a handbook~\cite{PZ},
 no other solution than Eq.~(\ref{usol}) is found yet.

\subsection{Planar (inflationary) coordinates $({\sf t},x^{i})$}
A noticed character of the pure dS${}_{d}$ from Eq.~(\ref{9}) is the fact 
that it is maximally symmetric. 
Suppose a comoving observer in the pure dS${}_{d}$,
then he or she may find maximally-symmetric spatial hypersurface orthogonal 
to his or her time direction. The corresponding planar (inflationary)
metric takes the form
\begin{equation}\label{plme}
ds^2=-d{\sf t}^2+a^2({\sf t}/l)\gamma_{ij}dx^idx^j\;,
\end{equation}
where $a({\sf t}/l)$ is cosmic scale factor and $(d-1)$-dimensional 
spatial metric $\gamma_{ij}$ of  
hypersurface should also carry maximal symmetries as a defining property :
\begin{equation}\label{d-1r}
{}^{d-1}R_{ijkl}=k (\gamma_{ik}\gamma_{jl}-\gamma_{il}\gamma_{jk})\;,
\end{equation}
where $\displaystyle{k=\frac{a^4}{(d-1)(d-2)}\; {}^{d-1}R}$. 

Note that every nonvanishing component of the $d$-dimensional Riemann 
curvature 
$R_{\mu\nu\rho\sigma}$ is expressed by the $(d-1)$-dimensional 
metric (\ref{plme}) such as
\begin{eqnarray}
R_{{\sf t}i{\sf t}j}&=& - l^2a\ddot{ a}\gamma_{ij}\;, \\
R_{{\sf t}ijk}&=& -a\dot{a}\left(\partial_i \gamma_{kj}
+\partial_j \gamma_{ki}-\partial_k\gamma_{ij}\right)\;,  \\
R_{ijkl}&=&(ka^2+a^2\dot{a}^2) 
(\gamma_{ik}\gamma_{jl}-\gamma_{il}\gamma_{jk})\;,
\end{eqnarray}
where the overdot denotes derivative of the rescaled time variable, 
$\dot{a}\equiv da({\sf t}/l)/d({\sf t}/l)$. 
So does those of the Ricci tensor
\begin{eqnarray}
R_{{\sf t}{\sf t}} &=& -\frac{d-1}{a}\ddot{a}\; ,\\
R_{{\sf t}i} &=&  \frac{\dot{a}}{a} \gamma^{jk} (\partial_j \gamma_{ik}
-\partial_i \gamma_{jk}+\partial_k \gamma_{ji})\; ,\\
R_{ij}&=& [a\ddot{a}+(d-2)(\dot{a}^2+k)]\gamma_{ij}\;, 
\end{eqnarray}
and that of the scalar curvature 
\begin{equation}\label{cspl}
\displaystyle {R= (d-1)\frac{2a\ddot{a}+(d-2) \dot{a}^2 + (d-2)k}{a^2}}.
\end{equation}
In the de Sitter spacetime of our interest, we only have a positive vacuum 
energy (or equivalently a positive cosmological constant) as the matter 
source given in Eq.~(\ref{sour}). Let us interpret it in terms of a 
cosmological perfect fluid of which energy-momentum tensor can be written
\begin{equation}
T_{\mu\nu}=(p+\rho)U_{\mu}U_{\nu}+pg_{\mu\nu},
\end{equation}
where $\rho$ and $p$ are the energy density and pressure respectively as 
measured in the rest frame, and $U_{\mu}$ is the velocity of the fluid.
Since the fluid is at rest for a comoving observer, the velocity of the fluid
$U_{\mu}$ is
\begin{equation}
U^{\mu}=(1,0,0,...,0),
\end{equation}
and then $T^{\mu}_{\;\nu}\equiv {\rm diag}(-\rho,p,\dots , p)$.
Therefore, the equation of state $w=p/\rho=-1$
tells it is a perfect fluid with a positive constant density 
and a negative constant pressure~:
\begin{equation}
\rho=-p=\frac{\Lambda}{8\pi G}.
\end{equation} 
Obviously it is also consistent with conservation of the energy-momentum
tensor (\ref{sour}), i.e., $\nabla_{\mu} T^{\mu}_{\; \nu}=0$.

Due to isotropy and homogeneity, spatial part of the metric (\ref{plme})
$\gamma_{ij}$ is rewritten by well-known Robertson-Walker metric 
in terms of $(d-1)$-dimensional 
spherical coordinates $({\sf r},\theta_{a})$, $(a=1,2,\cdots , d-2)$ :
\begin{equation}\label{plms}
ds^{2}=-d{\sf t}^{2}+a^{2}({\sf t}/l)
\left[\frac{d{\sf r}^{2}}{1-k({\sf r}/l)^{2}}+
{\sf r}^{2}d\Omega^{2}_{d-2}\right],
\end{equation}
where $k$ can have 0 (flat) or $-1$ (open) or $+1$ (closed). 
Under the metric (\ref{plms}), the Einstein equations (\ref{2}) are 
summarized by two Friedmann equations :
\begin{eqnarray}
\left(\frac{\dot{a}}{a}\right)^2 &=&
\frac{4\pi G}{d-2}\left[\frac{d}{d-1} \rho - (d-4)p\right]-\frac{k}{a^2}
=\frac{d-2}{2(d-1)}\Lambda-\frac{k}{a^{2}},
\label{frid1}\\
\frac{\ddot{a}}{a} &=& - 4\pi G
\left( \frac{\rho}{d-1} +p\right)
=\frac{d-2}{2(d-1)}\Lambda . 
\label{frid2}
\end{eqnarray}
The right-hand side of Eq.~(\ref{frid2}) is always positive in the dS${}_{d}$
of a positive cosmological constant. Thus, the universe depicted by
the dS${}_{d}$ is accelerating or equivalently deceleration parameter 
\begin{equation}\label{depa}
q=-\frac{a\ddot{a}}{a^2}
\end{equation} 
is observed to be negative. When $k=0$ or $-1$,
the right-hand side of Eq.~(\ref{frid1}) is always positive. It means that, 
once the universe started expanding, it is eternally expanding. Or 
equivalently, once Hubble parameter 
\begin{equation}\label{Hub}
H(t)\equiv \frac{\dot{a}}{a}
\end{equation}
 was observed to be positive, 
the rate of expansion of the universe always remains to be positive. 
 Even for $k=+1$ case, once the cosmic scale
factor $a({\sf t}/l)$ arrives at critical size $a_{\rm cr}$ such as
$a_{\rm cr}=\sqrt{2(d-1)/(d-2)\Lambda}$, and then the universe continues 
eternal expansion. The exact solutions of the Friedmann equations 
(\ref{frid1})--(\ref{frid2}) are nothing but inflationary solutions
consistent with the above arguments
\begin{equation}\label{plso}
a({\sf t}/l)
\left\{
\begin{array}{ll}
=l\sinh ({\sf t}/l), & {\rm for}~k=-1 \\
\propto \exp(\pm{\sf t}/l), & {\rm for}~k=0 \\
=l\cosh ({\sf t}/l), & {\rm for}~k=+1
\end{array}
\right. .
\end{equation}
If we interpret singularity at $a =0$ as the Big Bang of the 
creation of the universe, then open universe of $k=-1$ experienced it 
at ${\sf t}=0$ and an expanding flat universe of $k=0$ did it at past 
infinity $ {\sf t}=-\infty$ while closed universe of $k=+1$ did not.

\begin{figure}[ht]
\centerline{\psfig{figure=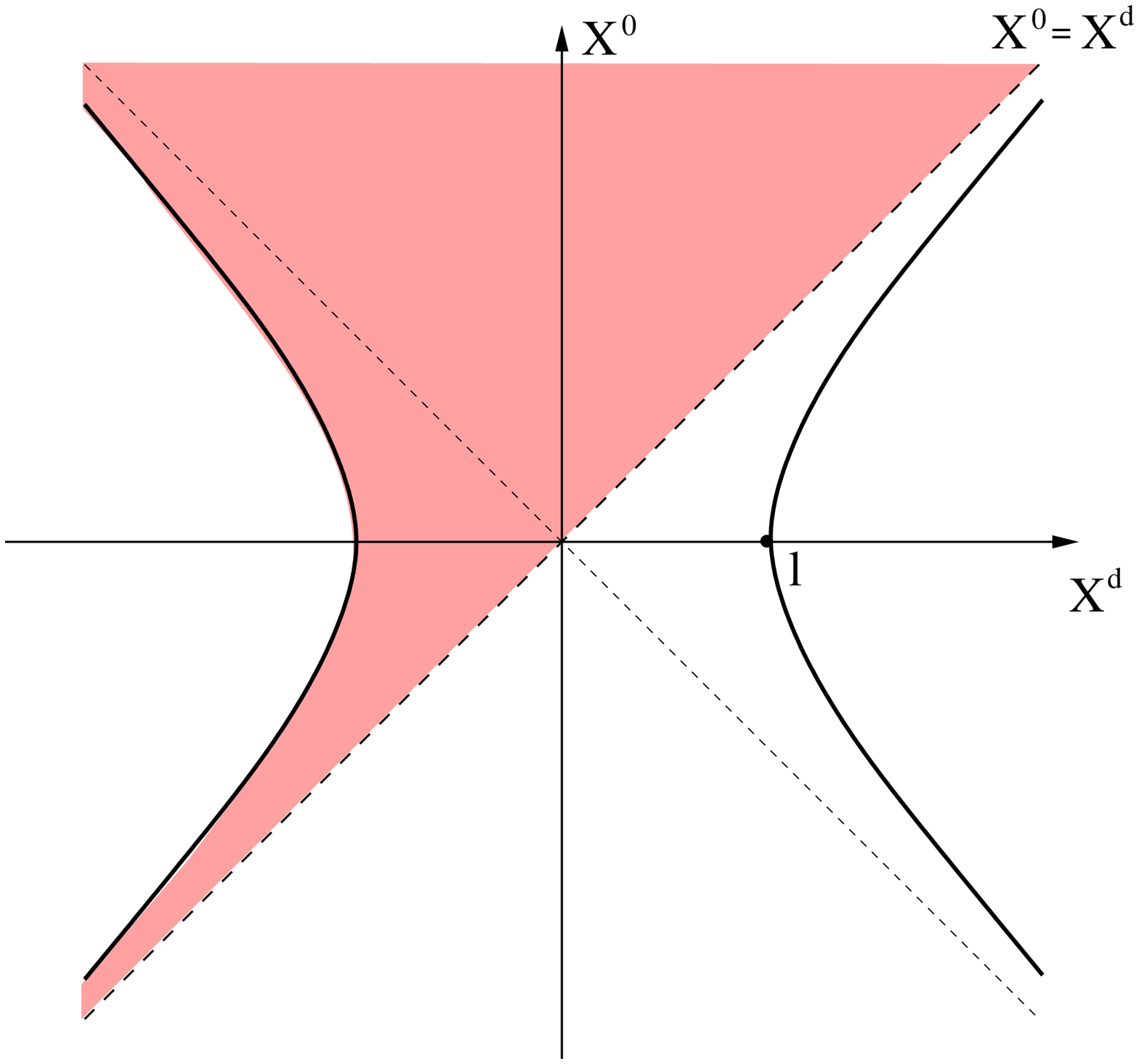,height=90mm}}
\caption{Upper-half the dS${}_{d}$ covered by our planar coordinates
is given by shaded region.}
\label{dsrfig2}
\end{figure}

The constraint of the hyperboloid (\ref{12}) embedded in $(d+1)$-dimensional
Minkowski spacetime can be decomposed into two constraints by introducing an
additional parameter $\sf t$, of which one is a 2-dimensional hyperbola of
radius $\sqrt{1-\left(\frac{x^i}{l}\right)^2 e^{2{\sf t}/l}}$
\begin{equation}\label{rrad}
-\left({\frac{X^0}{l}}\right)^2 +
\left({\frac{X^{d}}{l}}\right)^2 = 
1-\left(\frac{x^i}{l}\right)^2 e^{2{\sf t}/l},
\end{equation}
and the other is a $(d-1)$-dimensional sphere of radius
$\frac{x^i}{l}e^{{\sf t}/l}$
\begin{equation}\label{llet}
 \left({\frac{X^1}{l}}\right)^2 + \cdots
+ \left({\frac{X^{d-1}}{l}}\right)^2
= \left(\frac{x^i}{l}\right)^2 e^{2{\sf t}/l},
\end{equation}
where $(x^i)^2$ denotes the sum over the index $i$.
Therefore, a nice coordinate system to implement the above two constraints
(\ref{rrad})--(\ref{llet}) is
\begin{eqnarray}\label{scct}
\frac{X^0}{l} &=&- \sinh \frac{\sf t}{l}+\frac{(x^i/l)^2}{2}e^{{\sf t}/l},
\nonumber\\
\frac{X^i}{l} &=& \frac{x^i}{l}e^{{\sf t}/l}~~~~~(i=1, 2,\cdots,d-1),
\nonumber\\
\frac{X^d}{l} &=&- \cosh \frac{\sf t}{l}-\frac{(x^i/l)^2}{2}e^{{\sf t}/l},
\end{eqnarray}
where range of $x^i$ is $-\infty< x^i<\infty$ and that of ${\sf t}$
is $~-\infty< {\sf t}<\infty$. Since $-X^{0}+X^{d}=-l\exp(-{\sf t}/l)\le 0$,
our planar coordinates (\ref{scct}) cover only upper-half 
the dS${}_{d}$ as shown in
Fig.~\ref{dsrfig2}. Lower-half the dS${}_{d}$ can be described by changing 
the $(d+1)$-th coordinate $X^{d}$ to $X^d /l=\cosh ({\sf t}/l)+
(x^i)^{2} e^{{\sf t}}/2l^{2}$.
Inserting these transformations (\ref{scct}) into the flat $(d+1)$-dimensional
Minkowski metric (\ref{11}), we obtain a familiar form of the planar
metric of the dS${}_d$ :
\begin{equation}\label{plm}
ds^{2}= - d{\sf t}^{2} + e^{2{\sf t}/l}d{x^i}^2
\end{equation}
which coincides exactly with the flat solution (\ref{plso}) found by solving
the Einstein equations (\ref{frid1})--(\ref{frid2}).

Since the metric (\ref{plm}) is not isometric under time translation, 
$\partial/\partial {\sf t}$ is not a timelike Killing vector. This 
nonexistence implies no notion of conserved energy and Hamiltonian, 
which hinders 
description of quantum gravity in the planar coordinates.
However, the metric (\ref{plm}) is independent of  
$x^i$, so it satisfies form-invariance (\ref{fin}).
Therefore, $\partial /\partial x^i$'s are spacelike Killing vectors and
the spatial geometry involves translational symmetries and rotational 
symmetries. 


\subsection{Static coordinates $(t,r,\theta_{a}),~a=1,2,\cdots ,d-2$}
\begin{figure}[ht]
\centerline{\psfig{figure=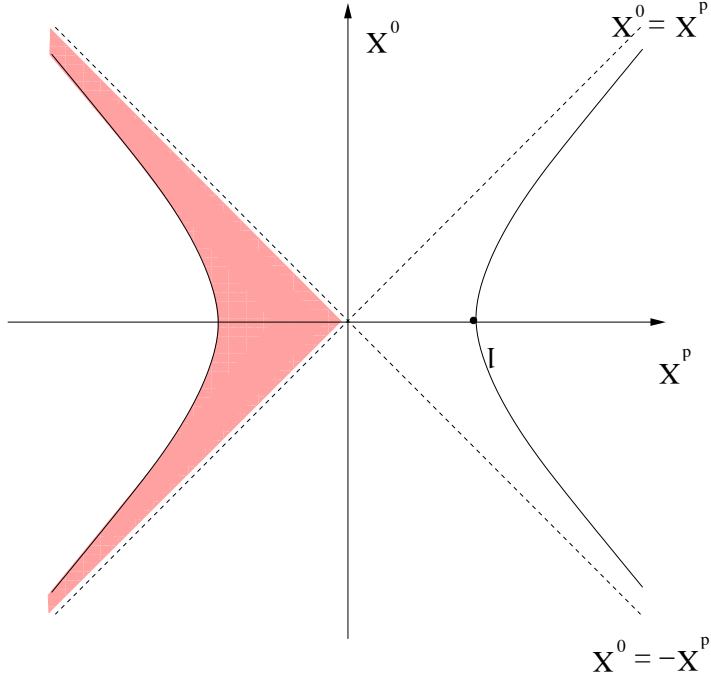,height=90mm}}
\caption{The region bounded by the de Sitter horizon $(r\le l)$ is shown by
 shaded region.}
\label{dsrfig3}
\end{figure}

The constraint of the hyperboloid (\ref{12}) embedded in $(d+1)$-dimensional
Minkowski spacetime is again decomposed into two constraints by introducing an
additional parameter $r$, of which one is a 2-dimensional hyperbola of
radius $\sqrt{1-\left(r/l\right)^2}$
\begin{equation}\label{radii}
-\left({\frac{X^0}{l}}\right)^2 +
\left({\frac{X^{d}}{l}}\right)^2 = 1-\left({\frac{r}{l}}\right)^2
\end{equation}
and the other is a $(d-1)$-dimensional sphere of radius $r/l$
\begin{equation}\label{let}
- \left({\frac{X^1}{l}}\right)^2 + \cdots
+ \left({\frac{X^{d-1}}{l}}\right)^2 = \left({\frac{r}{l}}\right)^2.
\end{equation}
Therefore, a nice coordinate system to implement the above two 
constraints
(\ref{radii})--(\ref{let}) is
\begin{eqnarray}\label{sct}
\frac{X^0}{l} &=& -\sqrt{1-\left( \frac{r}{l} \right)^2} \sinh \frac{t}{l},
\nonumber\\
\frac{X^i}{l} &=& \frac{r}{l} \omega ^i, ~~~~~(i=1, 2,\cdots,d-1),
\nonumber\\
\frac{X^d}{l} &=& -\sqrt{1-\left( \frac{r}{l} \right)^2} \cosh \frac{t}{l},
\end{eqnarray}
where $\omega^i$'s were given in Eq.~(\ref{omega}) and 
$r~(0\leq r<\infty)$ will be identified with radial coordinate 
of static coordinate system.
Since $-X^{0}+X^{d}=-\sqrt{l^{2}-r^{2}}\exp(-{\sf t}/l)\le 0$ and 
$X^{0}+X^{d}=-\sqrt{l^{2}-r^{2}}\exp({\sf t}/l)\le 0$, the region of
$r\le l$ covers only a quarter of the whole dS${}_{d}$ as shown by the 
shaded region in Fig.~\ref{dsrfig3}.
Later, $r=l$ will be identified by a horizon of the pure de Sitter
spacetime.

Inserting these transformations (\ref{sct}) into the 
$(d+1)$-dimensional Minkowski metric (\ref{11}), we obtain a 
familiar form of static metric of the dS${}_d$ :
\begin{equation}\label{pcs}
ds^{2}= - \left[ 1 - \left( \frac{r}{l}\right)^2 \right] dt^{2} +
\frac{dr^2}{\left[ 1 - \left( \frac{r}{l}\right)^2 \right]}
+ r^2 d \Omega^2 _{d-2},
\end{equation}
where 
\begin{eqnarray}
d\Omega^2_{d-2} & = & d\theta_1^2+\sin^2\theta_1d\theta_2^2+\cdots
+\sin^2\theta_1\cdots\sin^2\theta_{d-3}d\theta^2_{d-2} \nonumber\\
& = & \sum_{b=1}^{d-2}\left( \prod_{a=1}^{b-1} \sin^2 \theta_a \right) 
d\theta_b^2.
\nonumber
\end{eqnarray}
The metric (\ref{pcs}) is form-invariant (\ref{fin}) for both 
time translation and rotation of the coordinate $\theta_{d-2}$ so that
 we have two Killing vectors, $\partial / \partial t$ and $\partial/
\partial \theta_{d-2}$. Correspondingly, spacetime geometry has axial 
and time translational symmetries.
Therefore, Hamiltonian is well-defined in the static coordinates (\ref{pcs})
but unitarity is threatened by existence of the horizon at $r=l$.

In order to describe the system with rotational symmetry 
in $d$-dimensions, a static observer may introduce the static 
coordinate system where the metric involves two independent functions 
of the radial coordinate $r$, e.g.,
$\Omega(r)$ and $A(r)$ :
\begin{equation}\label{pc}
ds^{2}= -e^{2\Omega(r)} A(r)dt^{2} + \frac{dr^2}{A(r)}
+ r^2 d \Omega^2 _{d-2}.
\end{equation}
Curvature scalar of the metric (\ref{pc}) is computed as
\begin{equation}\label{csst}
R=(d-2)\left[\frac{(d-3)(1-A)}{r^2}
-\frac{2}{r}\left(\frac{dA}{dr}+A\frac{d\Omega}{dr}\right)
\right]
-\left[\frac{d^2A}{dr^2}+2A\frac{d^2\Omega}{dr^2}+2A\left(\frac{d\Omega}{dr}\right)^2+3\frac{dA}{dr}\frac{d\Omega}{dr}\right].
\end{equation}
From Eqs. (\ref{sust}) and (\ref{swst}),
simplified form of the Einstein equations (\ref{2}) is
\begin{equation}\label{dneq}
\frac{d-2}{r}\frac{d\Omega}{dr} = 0, 
\end{equation}
\begin{equation}\label{eneq}
\frac{d-2}{r^{d-2}}\frac{d}{dr}\left[r^{d-3}(1-A)\right] = 
 \frac{(d-1)(d-2)}{l^2} .
\end{equation}
Schwarzschild-de Sitter solution of Eqs.~(\ref{dneq}) and 
(\ref{eneq}) is
\begin{equation}
\Omega=\Omega_0 ~~\mbox{and}~~ A=1-\frac{r^2}{l^2}-
\frac{2GM}{r^{d-3}}.
\end{equation}
Here an integration constant $\Omega_0$ can always be absorbed by a scale 
transformation of the time variable $t$, $dt \rightarrow e^{-\Omega_0}dt$, 
and the other integration constant $M$ is chosen to be zero for the pure 
de Sitter spacetime of our interest, which is proportional to the mass 
of a Schwarzschild-de Sitter black hole~\cite{GH}.
Then the resultant metric coincides exactly with that of Eq.~(\ref{pcs}).

\setcounter{equation}{0}
\section{Geodesics}
Structure of a fixed curved spacetime is usually probed by classical motions 
of a test particle. The shortest curve, the geodesic, connecting two
points in the de Sitter space is determined by a minimum 
of its arc-length $\sigma$
for given initial point $P_{\rm i}$ and end-point $P_{\rm f}$, and is
parametrized by an arbitrary parameter $\lambda$ such as $x^{\mu}(\lambda)$ :
\begin{equation}\label{arc}
\sigma=\int_{P_{\rm i}}^{P_{\rm f}}d\sigma =
\int_{\lambda_{\rm i}}^{\lambda_{\rm f}}
d\lambda \,\frac{dI}{d\lambda}=\int_{\lambda_{\rm i}}^{\lambda_{\rm f}}
d\lambda \, L=\int_{\lambda_{\rm i}}^{\lambda_{\rm f}}d\lambda
\, \sqrt{g_{\mu\nu}\frac{dx^{\mu}}{d\lambda}\frac{dx^{\nu}}{d\lambda}}
=({\rm extremum}).
\end{equation}
According to the variational principle, the geodesic must obey 
second-order Euler-Lagrange equation 
\begin{equation}\label{geq}
\frac{d^{2}x^{\mu}}{d\lambda^{2}}+\Gamma^{\mu}_{\nu\rho}
\frac{dx^{\nu}}{d\lambda}\frac{dx^{\rho}}{d\lambda}=0.
\end{equation}
When the parameter $\lambda$ is chosen by the arc-length $\sigma$ itself, a 
force-free test particle moves on a geodesic.

In this section, we analyze precisely possible geodesics in the four
coordinate systems of the de Sitter spacetime, obtained in the previous 
section. We also introduce several useful quantities in each coordinate
system and explain some characters of the obtained geodesics.   
 
\subsection{Global (closed) coordinates}\label{glogeo}
Lagrangian for the geodesic motions (\ref{arc}) is read from the metric 
(\ref{gcs}) in the global coordinates~:
\begin{equation}
L^{2}=-
\left(\frac{d\tau}{d\lambda}\right)^2+l^2\cosh^2 \left( \frac{\tau}{l}\right)
\sum_{j=1}^{d-1}\left( \prod _{i=1}^{j-1} \sin ^2 \theta _{i}\right)
\left( \frac{d\theta_j}{d\lambda}\right)^2,
\end{equation}
where $\lambda$ is an affine parameter.
Corresponding geodesic equations (\ref{geq}) are given by $d$-coupled equations
\begin{equation}\label{gteq1}
\displaystyle
\frac{d^2\tau}{d\lambda^2}+l\sinh(\tau/l)\cosh(\tau/l) \prod_{j=1}^{i-1}\sin^2
\theta_j\left(\frac{d\theta_i}{d\lambda}\right)^2 = 0 \;,
\end{equation}
\begin{equation}\label{geq2}
\displaystyle
\frac{d^2\theta_i}{d\lambda^2}+ \frac{2}{l} \frac{\sinh(\tau/l)}{\cosh(\tau/l)}
 \frac{d\tau}{d\lambda}
\frac{d\theta_i}{d\lambda}-\sin\theta_i\cos\theta_i \prod_{j=i+1}^{j-1}\sin^2
\theta_j\left(\frac{d\theta_j}{d\lambda}\right)^2+2\left(\sum_{k=1}^{i-1}
\frac{\cos\theta_k}{\sin\theta_k} \frac{d\theta_k}{d\lambda}\right) 
\frac{d\theta_j}{d\lambda} = 0 \;.
\end{equation}

\begin{figure}[ht]
\centerline{\psfig{figure=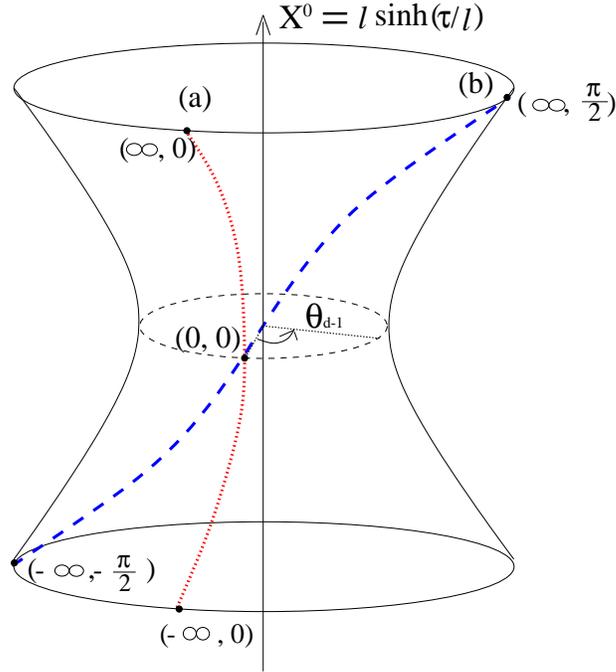,height=90mm}}
\caption{Two representative geodesic motions in the global coordinates :
(a) dotted line for $J=0$, (b) dashed line for $J\ne 0$.}
\label{geo1}
\end{figure}

Since $\theta_{d-1}$ is cyclic, $d\theta_{d-1}/d\lambda$ in Eq.~(\ref{geq2}) 
is replaced by a constant of motion $J$ such as 
\begin{equation}\label{am} 
l^2\cosh^2(\tau/l)\prod_{j=1}^{i-1}\sin^2
\theta_j \frac{d\theta_{d-1}}{d\lambda}=J. 
\end{equation}
Let us recall a well-known fact that any geodesic connecting arbitrary 
two points on a $(d-1)$-dimensional sphere should be located on its 
greatest circle. Due to rotational symmetry on the $S^{d-1}$, 
one can always orient the coordinate system so that the radial 
projection of the orbit coincides with the equator,
\begin{equation}\label{fixa}
\theta_1=\theta_2=\theta_3=\cdots=\theta_{d-2}=\frac{\pi}{2}, 
\end{equation}
of the spherical coordinates. This can also be confirmed by an explicit 
check that Eq.~(\ref{fixa}) should be a solution of Eq.~(\ref{geq2}).
It means that a test particle has at start and continues 
to have zero momenta in the $\theta_i$-directions ($i=1, 2,\cdots , d-2$).
Therefore, the system of our interest reduces from $d$-dimensions to 
($1+1$)-dimensions without loss of generality.
Insertion of Eq.~(\ref{fixa}) into Eq.~(\ref{am}) gives
\begin{equation}\label{fixb}
\frac{d\theta_{d-1}}{d\lambda}=\frac{J}{l^2\cosh^2(\tau/l)},
\end{equation} 
so that the remaining equation~(\ref{gteq1}) becomes
\begin{equation}\label{taud}
\frac{d^2\tau}{d\lambda^2}=-\frac{J^2}{l^3}
\frac{\sinh(\tau/l)}{\cosh^3(\tau/l)}.
\end{equation}
Integration of Eq.~(\ref{taud}) arrives at the conservation of 
\lq energy\rq~$E$
\begin{equation}\label{ener}
E= \frac{1}{2}\left(\frac{d\tau}{d\lambda}\right)^2 + V_{\rm eff}(\tau/l),
\end{equation}
where the effective potential $V_{\rm eff}$ is given by
\begin{equation} 
V_{\rm eff}(\tau/l) 
= \left\{ 
\begin{array}{cl} 
 0\;\;\; & (J=0) \\
 -\frac{1}{2}\left(\frac{J}{l}\right)^2 \frac{1}{\cosh^2(\tau/l)} 
\;\;\; & (J \neq 0)
\end{array}
, \right.
\end{equation}
and thereby another constant of motion $E$ should be bounded below, i.e.,
$E\ge -\frac{1}{2}(J/2l)^2$. Eliminating the affine parameter 
$\lambda$ in both Eq.~(\ref{fixb}) and Eq.~(\ref{ener}), we obtain the 
orbit equation which is integrated as an algebraic equation : 
\begin{equation}
\tan\theta_{d-1}=\frac{J}{l}\frac{\sinh{(\tau/l)}}
{\sqrt{(J/l)^2+2E\cosh^2(\tau/l)}}.
\end{equation}
Since the spatial sections are sphere $S^{d-1}$ of a constant positive 
curvature and Cauchy surfaces, their geodesic normals of $J=0$ are lines
which monotonically contract to a minimum spatial separation and then
re-expand to infinity (See dotted line in Fig.~\ref{geo1}). 
Another representative geodesic motion of $J\ne 0$ 
(dashed line) is also sketched in Fig.~\ref{geo1}.
For example, suppose that the affine parameter $\lambda$ is identified with
 the coordinate time $\tau$. Then, we easily confirm from the above analysis
 that every geodesic emanating from any point can be extended to infinite 
 values of the affine parameters in both directions, $\lambda = \tau \in
(-\infty, \infty)$, so that the de Sitter spacetime is said to be 
geodesically complete. However, there exist spatially-separated points which
cannot be joined by one geodesic.

\subsection{Conformal coordinates}
Similar to the procedure in the previous subsection, we read 
automatically Lagrangian for geodesics (\ref{arc}) from the conformal  
metric (\ref{cc})~:
\begin{equation}
L^{2}
=\sec^2 \left( \frac{T}{l} \right) \left[ -\left( \frac{dT}{d\lambda } \right)^2
+l^2 \sum_{j=1}^{d-1} \left( \prod _{i=1}^{j-1} \sin ^2 \theta _{i}\right)
\left( \frac{d\theta_j}{d\lambda}\right)^2 \right],
\end{equation}
and then corresponding geodesic equations (\ref{geq}) are
\begin{equation}\label{coge1}
\displaystyle
 \frac{d^2T}{d\lambda^2}+\frac{1}{l}\frac{\sin(T/l)}{\cos(T/l)}
\left(\frac{dT}{d\lambda}\right)^2
+l\frac{\sin(T/l)}{\cos(T/l)} \prod_{j=1}^{i-1}\sin^2
\theta_j \left(\frac{d\theta_i}{d\lambda}\right)^2 = 0 \;,
\end{equation}
\begin{equation}\label{coge2}
\displaystyle 
\frac{d^2\theta_i}{du^2}+\frac{2}{l}\frac{\sin(T/l)}{\cos(T/l)}
\frac{dT}{d\lambda} \frac{d\theta_i}{d\lambda}
-\sin\theta_i\cos\theta_i \prod_{j=i+1}^{j-1}\sin^2\theta_j
\left(\frac{d\theta_j}{d\lambda}\right)^2+2\left(\sum_{k=1}^{i-1}
\frac{\cos\theta_k}{\sin\theta_k} \frac{d\theta_k}{d\lambda}\right) 
\frac{d\theta_j}{d\lambda} = 0 
\;. 
\end{equation}

\begin{figure}[ht]
\centerline{\psfig{figure=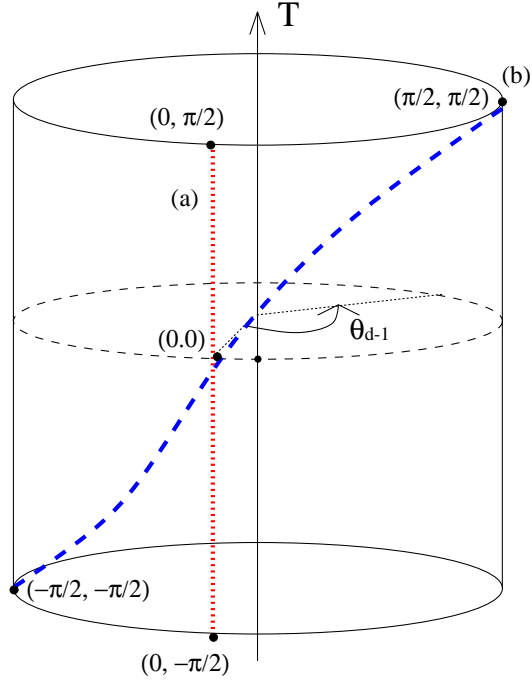,height=90mm}}
\caption{Geodesics on the $(T,\theta_{d-1})$ plane : (a) dotted line 
for $J=0$, (b) dashed line for $J\neq 0$.}
\label{gd11}
\end{figure}

Since $\theta_{d-1}$ is cyclic, $d\theta_{d-1}/d\lambda$ is replaced 
by a constant
of motion $J$ such as 
\begin{equation}\label{am2} 
l^2 \cos^{-2}(T/l) \prod_{j=1}^{i-1}\sin^2
\theta_j \frac{\theta_{d-1}}{d\lambda}=J. 
\end{equation}
Since the $(d-1)$-angular coordinates $\{\theta_i\}$ constitute a 
($d-1$)-dimensional sphere, the same argument around Eq.~(\ref{fixa})
 is applied and thereby the system of our interest  
 reduces again from $d$-dimensions to $(1+1)$-dimensions  
without loss of generality.
Substituting of Eq.~(\ref{fixa})  into Eq.~(\ref{am2}), we have
\begin{equation}\label{fixb2}
\frac{d\theta_{d-1}}{d\lambda}=\frac{J}{l^2}\cos^2(T/l)
\end{equation} 
and rewrite Eq.~(\ref{coge1}) as
\begin{equation}\label{taud2}
\frac{d^2T}{d\lambda^2}+\frac{1}{l}\tan(T/l)\left(\frac{dT}{d\lambda}
\right)^2 +l\tan(T/l)\left(\frac{d\theta_{d-1}}{d\lambda}\right)^2= 0\;.
\end{equation}
Eq.~(\ref{taud2}) is integrated out and we obtain another conserved 
quantity $E$ : 
\begin{equation}\label{ener2}
E\equiv-\frac{l^2}{2}\left(\frac{d\theta_{d-1}}{d\lambda}\right)^2
 = \frac{1}{2}\left(\frac{dT}{d\lambda}\right)^2-\cos^2(T/l)\;.
\end{equation}
In Eq.~(\ref{ener2}) we rescaled the affine parameter $\lambda$ 
in order to absorb a redundant constant.
Combining Eq.~(\ref{fixb2}) and Eq.~(\ref{ener2}), we obtain an orbit 
equation  expressed by elliptic functions :
\begin{equation}
\theta_{d-1} = \frac{J}{l\sqrt{2}} \left[ \sqrt{1+E} 
 \, EllipticE\left(\frac{T}{l},\frac{1}{E+1}\right) 
- \frac{E}{ \sqrt{1+ E}}\,
EllipticF\left(\frac{T}{l},\frac{1}{E+1}\right) 
\right] \;,
\end{equation}
where
\begin{equation}
EllipticE\left(\frac{T}{l},\frac{1}{E+1}\right)=
\int dT \sqrt{1-\frac{1}{E+1}\sin^2(T/l)}\;,  
\end{equation}
\begin{equation}
  EllipticF\left( \frac{T}{l},\frac{1}{E+1}\right) =
\int dT \frac{1}{\sqrt{1-\frac{1}{E+1}\sin^2(T/l)}}\;.
\end{equation}
($T, \theta_{d-1}$) represents a cylinder of finite height as shown in
Fig.~\ref{gd11}. Geodesic of zero energy, $E=0$, (or equivalently
zero angular momentum, $J=0$), is shown as a dotted line, and that of
positive energy, $E>0$, (or nonvanishing angular momentum, $J\neq0$),
as a dashed line in Fig.~\ref{gd11}.

\subsection{Planar (inflationary) coordinates}\label{pigo}
Lagrangian for the geodesic motion (\ref{arc}) for the flat space of $k=0$
is read from the metric (\ref{plms}) :
\begin{equation}
L^{2}
=-\left(\frac{d{\sf t}}{d\lambda}\right)^2+e^{2{\sf t}/l}
\left(\frac{d{\sf r}}{d\lambda}\right)^2
+e^{2{\sf t}/l}{\sf r}^2 
\sum_{b=1}^{d-2} \left( \prod _{a=1}^{b-1} \sin ^2 \theta _{a}\right)
\left( \frac{d\theta_b}{d\lambda}\right)^2 ,
\end{equation}
and the corresponding geodesic equations (\ref{geq}) are
\begin{equation}\label{plge1}
\frac{d^2\sf t}{d\lambda^2}
+\frac{1}{l}e^{2{\sf t}/l}
\left(\frac{d{\sf r}}{d\lambda}\right)^2+\frac{e^{2{\sf t}/l}}{l}
{\sf r}^2\sum_{b=1}^{d-2} \left( \prod _{a=1}^{b-1} \sin ^2 \theta _{a}\right)
\left( \frac{d\theta_b}{d\lambda}\right)^2
 = 0 \;,
\end{equation}
\begin{equation}\label{plge2}
\frac{d^2{\sf r}}{d\lambda^2}+\frac{2}{l} \frac{d\sf t}{d\lambda} 
\frac{d{\sf r}}{d\lambda}
-{\sf r} \sum_{b=1}^{d-2} \left( \prod _{a=1}^{b-1} \sin ^2 \theta _{a}\right)
\left( \frac{d\theta_b}{d\lambda}\right)^2 =0 \;,
\end{equation}
\begin{equation}\label{plge3}
\frac{d^2\theta_a}{d\lambda^2}+\frac{2}{l}
\frac{d{\sf t}}{d\lambda} \frac{d\theta_a}{d\lambda}
+\frac{2}{{\sf r}}
\frac{d{\sf r}}{d\lambda} \frac{d\theta_a}{d\lambda}
-\sin\theta_a\cos\theta_a \prod_{b=a+1}^{b-1}\sin^2\theta_b
\left(\frac{d\theta_b}{d\lambda}\right)^2+2\left(\sum_{c=1}^{a-1}
\frac{\cos\theta_c}{\sin\theta_c} \frac{d\theta_c}{d\lambda}\right) 
\frac{d\theta_b}{d\lambda} = 0 
\;.  
\end{equation}

Since $\theta_{d-2}$ is cyclic, $d\theta_{d-2}/d\lambda$ is replaced 
by a constant of motion $J$ such as 
\begin{equation}\label{am3} 
2e^{2{\sf t}/l}{\sf r}^2\prod_{b=1}^{d-3}\sin^2
\theta_b \frac{d\theta_{d-2}}{d\lambda}=J. 
\end{equation}
If we use a well-known fact that any geodesic connecting arbitrary 
two points on a $(d-2)$-dimensional sphere 
should be located on its greatest circle, 
one may choose without loss of generality 
\begin{equation}\label{fixa3}
\theta_1=\theta_2=\theta_3=\cdots=\theta_{d-3}=\frac{\pi}{2}, 
\end{equation}
which should be a solution of Eq.~(\ref{plge3}) due to rotational symmetry
on the $(d-2)$-dimensional sphere $S^{d-2}$. 
Therefore, the system of our interest  
reduces from $d$-dimensions to $(2+1)$-dimensions  
without loss of generality.\\
Insertion of Eq.~(\ref{fixa3})  into Eq.~(\ref{am3}) gives
\begin{equation}\label{fixb3}
\frac{d\theta_{d-2}}{d\lambda}=\frac{J}{2e^{2{\sf t}/l}{\sf r}^2},
\end{equation} 
so that the equations (\ref{plge1})--(\ref{plge2}) become 
\begin{equation}\label{taud3}
\frac{d^2\sf t}{d\lambda^2}
+\frac{e^{2{\sf t}/l}}{l}
\left(\frac{d{\sf r}}{d\lambda}\right)^2+\frac{J^2}{4l}
\frac{e^{-2{\sf t}/l}}{{\sf r}^2} =  0\;,
\end{equation}
\begin{equation}\label{taudd3}
\frac{d^2{\sf r}}{d\lambda^2}+\frac{2}{l} \frac{d\sf t}{d\lambda} 
\frac{d{\sf r}}{d\lambda}
-\frac{J^2}{4}\frac{e^{-4{\sf t}/l}}{{\sf r}^3} = 0\;.
\end{equation}
Eliminating the third term in both Eqs.~(\ref{taud3})--(\ref{taudd3})
by subtraction, we have the combined equation 
\begin{equation}\label{ener3}
{\sf r}\frac{d{\sf r}}{d\lambda}e^{2{\sf t}/l}=-l\frac{d{\sf t}}{d\lambda}+C,
\end{equation}
where $C$ is constant. 
Finally, for $C=0$, the radial equation (\ref{ener3}) is solved as  
\begin{equation}
{\sf r}=le^{{\sf -t}/l}.
\end{equation}
For $J=0$,  Eq.~(\ref{taudd3}) becomes
\begin{equation}
\frac{d}{d\lambda}\left(2e^{2{\sf t}/l}\frac{d{\sf r}}{d\lambda}\right)=0.
\end{equation}  
It reduces $dr/d\lambda =C^\prime e^{-2{\sf t}/l}$ and $C^\prime$ is constant.

Information on the scale factor $a({\sf t})$ is largely gained through 
the observation of shifts in wavelength of light emitted by distant sources.
It is conventionally gauged in terms of redshift parameter $z$ between 
two events, defined as the fractional change in wavelength~:
\begin{equation}\label{red}
z\equiv \frac{\lambda_{0}-\lambda_{1}}{\lambda_{1}},
\end{equation}
where $\lambda_{0}$ is the wavelength observed by us here after long journey
and $\lambda_{1}$ is that emitted by a distant source. 
For convenience,
we place ourselves at the origin ${\sf r}=0$ of coordinates since our de Sitter 
space is homogeneous and isotropic, and consider a photon traveling to us
along the radial direction with fixed $\theta_{a}$'s.
Suppose that a light is emitted from the source at time ${\sf t}_{1}$
and arrives at us at time ${\sf t}_{0}$. Then, from the metric~(\ref{plms}),
the null geodesic connecting (${\sf t}_{1}, {\sf r}_{1},\theta_1,\cdots,
\theta_{d-2} $) and (${\sf t}_{0}, 0, \theta_1,\cdots,\theta_{d-2} $)
relates coordinate time and distance as follows 
\begin{equation}\label{nur}
\int^{{\sf t}_0}_{{\sf t}_{1}}\frac{d{\sf t}}{a(\sf t)}=\int^{{\sf r}_{1}}_{0}
\frac{d{\sf r}}{\sqrt{1-k{\sf r}^2}}\equiv f({\sf r}_{1}).
\end{equation}
If next wave crest leaves ${\sf r}_{1}$ at time 
${\sf t}_{1}+\delta {\sf t}_{1}$ and reach us at time 
${\sf t}_{0}+\delta {\sf t}_{0}$, the time independence of $f({\sf r}_{1})$
provides
\begin{equation}\label{tim}
\int^{{\sf t}_1+\delta{\sf t}_1}_{{\sf t}_1} \frac{d{\sf t}}{a({\sf t})}
=\int^{{\sf t}_0+\delta{\sf t}_0}_{{\sf t}_0} \frac{d{\sf t}}{a({\sf t})}.
\end{equation}
For sufficiently short time $\delta {\sf t}_0$ (or 
$\delta {\sf t}_1$) 
the scale factor $a({\sf t})$ is approximated by a constant over the 
integration time in Eq.~(\ref{tim}), and, with the help of 
$\lambda_0=\delta {\sf t}_0 \ll |{\sf t}_1-{\sf t}_0|$ (or 
$\lambda_1=\delta {\sf t}_1 \ll |{\sf t}_1-{\sf t}_0|$), 
Eq.~(\ref{tim}) results in
\begin{eqnarray}\label{fac}
\frac{\lambda_{1}}{\lambda_{0}}&=&\frac{a({\sf t}_{1})}{a({\sf t}_{0})}\\
&=&1+H_{0}({\sf t}_{1}-{\sf t}_{0})
-\frac{1}{2}q_{0}H^{2}_{0}({\sf t}_{1}-{\sf t}_{0})^{2}+\cdots\, ,
\label{h0}
\end{eqnarray}
where $H_0=H({\sf t}_0)$ from Eq.~(\ref{Hub}) and $q_0=q({\sf t}_0)$ from
 Eq~(\ref{depa}).
Substituting Eq.~(\ref{fac}) into Eq.~(\ref{red}), we have
\begin{equation}\label{zz}
z=\frac{a({\sf t}_0)}{a({\sf t}_1)}-1.
\end{equation}
On the other hand, the comoving distance ${\sf r}_1$ is not measurable 
so that we can define the luminosity distance $d_L$~:
\begin{equation}\label{ldis}
d_L^2 \equiv \frac{L}{4\pi F},
\end{equation}
where $L$ is absolute luminosity of the source and $F$ is flux
 measured by the observer.
It is motivated from the fact that the measured flux $F$ is simply equal to
luminosity times one over the area around a source at distance $d$ in flat
 space.
In expanding universe, the flux will be diluted by the redshift of the light
 by a factor $(1+z)$ and the difference between emitting time and measured 
time. 
When comoving distance between the observer and the light source is 
${\sf r}_1$, a physical distance $d$ becomes $a_0r_1$,
 where $a_0$ is scale factor when the light is observed. 
Therefore, we have  
\begin{equation}\label{fl}
\frac{F}{L}= \frac{1}{4\pi a_0^2 r_1^2(1+z)^2}.
\end{equation}
 Inserting Eq.~(\ref{fl}) into Eq.~(\ref{ldis}), we obtain
\begin{equation}\label{dl}
d_L = a_0r_1(1+z).
\end{equation}
Using the expansion (\ref{h0}), Eq.~(\ref{zz}) is expressed by 
\begin{equation}\label{za}
\frac{1}{1+z}=1+H_{0}({\sf t}_{1}-{\sf t}_{0})
-\frac{1}{2}q_{0}H^{2}_{0}({\sf t}_{1}-{\sf t}_{0})^{2}+\cdots .
\end{equation}
For small $H_0({\sf t}_1-{\sf t}_0)$, Eq.~(\ref{za}) can be inverted to
\begin{equation}\label{zb}
{\sf t}_0-{\sf t}_1 = H_0^{-1}\left[z-\left(1+\frac{q_0}{2}\right)z^2+
\cdots\right].
\end{equation}
When $k=0$, the right-hand side of Eq.~(\ref{nur}) yields ${\sf r}$ and
 expansion of the left-hand side for small $H_0({\sf t}_1-{\sf t}_0)$ gives
\begin{eqnarray}
{\sf r}_1 &=& \int^{{\sf t}_0}_{{\sf t}_1}d{\sf t}
\left( \frac{1}{a({\sf t}_1)}-\frac{\dot{a}({\sf t}_1)}{a^2({\sf t}_1)}{\sf t}
+\cdots \right) \nonumber\\
&=& \frac{1}{a_0}\left[({\sf t}_0-{\sf t}_1) 
+ \frac{1}{2}H_0({\sf t}_0-{\sf t}_1)^2 +\cdots \right] \nonumber\\
&=& \frac{1}{a_0H_0}\left[z-\frac{1}{2}(1+q_0)z^2+\cdots\right],\label{rz}
\end{eqnarray}
where $a({\sf t}_1) \approx a_0+\dot{a}_0({\sf t}_1-{\sf t}_0)
+ \frac{1}{2} \ddot{a}_0({\sf t}_1-{\sf t}_0)^2+
\cdots $ was used in 
the second line and Eq.~(\ref{zb}) was inserted in the third line.
Replacing ${\sf r}_1$ in Eq.~(\ref{dl}) by Eq.~(\ref{rz}), 
we finally have Hubble's law~:
\begin{equation}\label{hl}
d_L=H_0^{-1}\left[z+\frac{1}{2}(1-q_0)z^2+\cdots\right]
\end{equation}
which relates the distance to a source with its observed red shift.
Note that we can determine present Hubble parameter $H_0$ and present 
deceleration parameter $q_0$ by measurement of the luminosity
 distances and redshifts.
As a reference, observed value of $H_0$ at present is $H_0 = 100 {\rm 
h\,km/s\,Mpc}$~($0.62 \stackrel{<}{_\sim} h \stackrel{<}{_\sim} 0.82$) 
so that corresponding time scale 
$T_{\rm universe}\equiv H_0^{-1}$ is about one billion year and length
 scale $ L_{\rm universe}\equiv c H_0^{-1}$ is about several thousand 
${\rm Mpc}$.
Since the scale factor is given by an exponential function, $a({\sf t})=
e^{{\sf t}/l}$, for the flat spacetime of $k=0$, the left-hand side of 
Eq.~(\ref{nur}) is integrated in a closed form~:
\begin{equation}
\frac{{\sf r}_1}{l}=e^{-{\sf t}_1/l}-e^{-{\sf t}_0/l}.
\end{equation}
In addition, the redshift parameter $z$ is expressed as 
\begin{equation}
z=e^{({\sf t}_0-{\sf t}_1 )/l}-1,
\end{equation}
so the luminosity distance $d_L$ in Eq.~(\ref{dl}) becomes
\begin{equation}\label{dl2}
d_L=l(z+z^2).
\end{equation}

Comparing Eq.~(\ref{dl2}) with the Hubble's law (\ref{hl}),
we finally confirm that the expanding flat space solution of $k=0$ 
has present Hubble parameter 
$H_0=1/l$ and the de Sitter universe is accelerating with present 
deceleration parameter $q_0=-1$ as expected.

\subsection{Static coordinates}
Geodesic motions parametrized by proper time $\sigma$ are described by the 
following Lagrangian read from the static metric (\ref{pcs}) :
\begin{equation}\label{sclag}
L^{2}
=-\left[1-\left(\frac{r}{l}\right)^2\right]\left(\frac{dt}{d\sigma}\right)^2
+\frac{1}{1-\left(\frac{r}{l}\right)^2} \left( \frac{dr}{d\sigma}\right)^2
+r^2 \sum_{b=1}^{d-2}\left( \prod _{a=1}^{b-1} \sin ^2 \theta _{a}\right)
\left( \frac{d\theta_b}{d\sigma}\right)^2.
\end{equation}
Since the time $t$ and angle $\theta_{d-2}$ coordinates are 
cyclic, the conjugate momenta $E$ and $J$ are conserved :
\begin{equation}\label{am4} 
J\equiv \frac{\partial L^{2}}{\partial \left( \frac{d \theta_{d-2} }{d\sigma} 
\right)}
=r^2\prod_{j=1}^{d-3}\sin^2
\theta_j \frac{d\theta_{d-2}}{d\sigma} \;, 
\end{equation}
\begin{equation}\label{como4}
\sqrt{-2E} \equiv \frac{\partial L^{2}}{\partial \left( \frac{d t}{d\sigma} 
\right)}  
=-\left[1-\left(\frac{r}{l}\right)^2\right]\frac{dt}{d\sigma}\;.
\end{equation}  
By the same argument in the subsection~\ref{pigo}, 
the system of our interest reduces from $d$-dimensions to 
($2+1$)-dimensions without loss of generality, and then Eq.~(\ref{am4}) 
becomes
\begin{equation}\label{fixb4}
\frac{d\theta_{d-2}}{d\sigma}=\frac{J}{r^2}\;.
\end{equation} 
From here on let us use rescaled variables $e=E/l^2$, $x=r/l$, and 
$j=J/l^2$. 
Note that $j$ cannot exceed $1$ due to the limitation of light velocity.
Then second-order radial geodesic equation from 
the Lagrangian~(\ref{sclag}) becomes 
\begin{equation}\label{taud4}
\frac{d^2x}{d\sigma^2}+\frac{x}{1
-x^2}\left(\frac{dx}{d\sigma}\right)^2
+2e\frac{x}{1-x^2}
-j^2\frac{1-x^2}{x^3}=0.
\end{equation}
Integration of Eq.~(\ref{taud4}) leads to conservation of the energy $e$
\begin{equation}\label{ener4}
e = \frac{1}{2}\left(\frac{dx}{d\sigma}\right)^2+\frac{V_{\rm eff}}{l^2} 
\leq0 \;,
\end{equation}
where the effective potential $V_{\rm eff}$ is given by
\begin{equation}\label{efpo}
V_{\rm eff}(x)= 
\frac{l^2}{2}\left(1-x^2\right)
\left(\frac{j^2}{x^2}-1 \right)\;. 
\end{equation}

\begin{figure}[ht]                                                   
\centerline{\psfig{figure=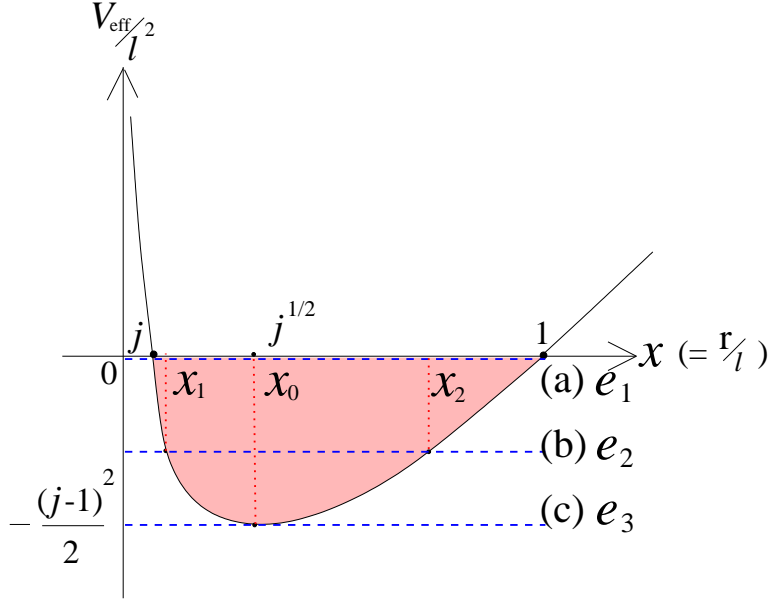,height=80mm}}                      
\caption{Effective potential and various available values of the energy $e$~: 
(a) $e_1=0$, 
(b) $-\frac{(j-1)^2}{2}<e_2<0$, (c) $e_3=-\frac{(j-1)^2}{2}$.}                  \label{geo4}                                                         
\end{figure}

As shown in Fig.~\ref{geo4}, the motion of a particle having the energy
$e_1$ can never lower than $x=j$ due to repulsive centrifugal force,
which becomes $0$ as $j$ approaches $0$.
The ranges except $j<r<1$ are forbidden by the fact that kinetic energy 
should be positive.    
For $e=e_2$ indicated in Fig.~\ref{geo4}, a test particle moves bounded orbit 
within two turning points $x_1$ and $x_2$. 
The perihelion $x_1$ and the aphelion $x_2$ were obtained from       
$dx/d\sigma=0$                                         
\begin{eqnarray}\label{semi}                                         
x_1^2 &=& \frac{j^2+1+2e-\sqrt{(j^2+1+2e)^2-4j^2}}{2},\nonumber\\    
x_2^2 &=& \frac{j^2+1+2e+\sqrt{(j^2+1+2e)^2-4j^2}}{2}.               
\end{eqnarray}                                                       
If the energy $e$ have minimum value of the effective potential $e=e_3$,
then the motion is possible only at $x=x_0$, so that the orbital motion 
should be circular.
Shaded region                                                    
in Fig.~\ref{geo4} is the allowable region for static              
observer, which is bounded by de Sitter horizon ($x=1$).                     
It corresponds to one                                                
fourth of the global hyperbolic region in Fig.~\ref{pen1}.         

Solving for $dr/d\sigma$ from Eq.~(\ref{ener4}), we have
\begin{equation}\label{geint}
d\sigma=\frac{dx^2}{2\sqrt{-x^4+(j^2+2e+1)x^2-j^2}}.
\end{equation}
Then the integration of both sides gives          
\begin{equation}\label{gesig}
x^2=\frac{j^2+1+2e}{2}+\frac{1}{2}
\sqrt{(j^2+1+2e)^2-4j^2} \sin{(2\sigma+C)},
\end{equation}
where $C$ is an integration constant. 
If we choose the perihelion $x_1$ given in Eq.~(\ref{semi}) for 
$\sigma=0$, $C$ is fixed by $-\pi/2$ and then the aphelion $x_2$ 
is given at $\sigma=\pi/2$. When the energy $e$ takes maximum value $e=0$,
position of the aphelion at  $x=1$ is nothing but the de Sitter horizon. 
So the elapsed proper time for the motion from the perihelion to the 
aphelion, ($-\pi/2,\pi/2$), is finite.

 Changing the proper time $\sigma$ in Eq.~(\ref{geint}) to coordinate
 time of a static observer ${\sf t}$ by means of Eq.~(\ref{como4}), we find 
\begin{equation}\label{getime}
dt=dx^2\frac{\sqrt{-2e}}{(x^2-1)\sqrt{-x^4+(j^2+2e+1)x^2-j^2}},
\end{equation}
and the integration for $e=e_2$ gives
\begin{equation}\label{dstime}
\tan[2(t-t_0)]=\frac{4e-(j^2-1+2e)(1-x^2)}{\sqrt{-2e}(x^2-1)
\sqrt{2e+(1-x^2)(1-j^2/x^2)}}.
\end{equation}
Note that the ranges are
\begin{equation}
j<x<1,\;\;\;0<j<1,\;\;\;~$$ and $$~-\frac{(j-1)^2}{2} < e_2 < 0.
\end{equation}

\begin{figure}[ht]
\centerline{\psfig{figure=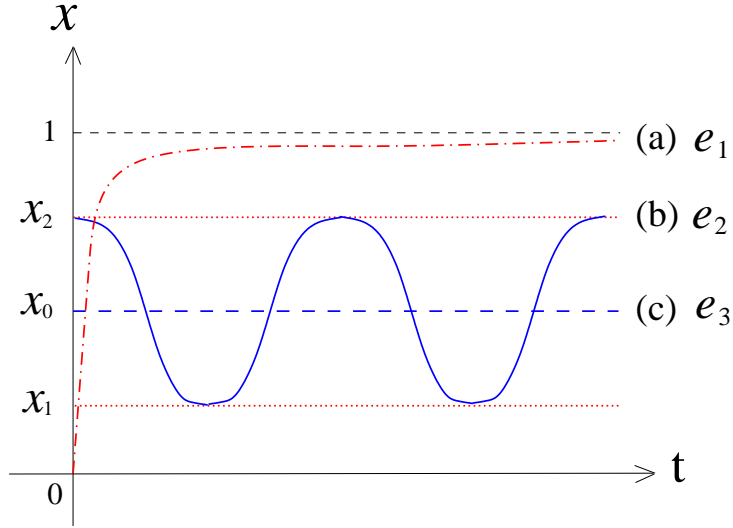,height=70mm}}
\caption{Radial geodesics on time $t$ : (a) dash-dot line for $e_1$, 
(b) solid line for $e_2$, (c) dashed line for $e_3$.} 
\label{radig}
\end{figure}

In the case of $e=e_1$, the integration of Eq.~(\ref{getime}) gives 
\begin{equation}\label{enzer}
t-t_0=\lim_{e\rightarrow 0}l\sqrt{-2e}
\frac{\sqrt{(1-j^2)(1-x^2)-(1-x^2)^2}}{(1-j^2)(x^2-1)}.
\end{equation}
To reach the de Sitter horizon at $x=1$, the energy $e$ should have a value
$e_1$ irrespective of the 
value of $j$ as shown in Fig.~\ref{radig}.
For $e=e_3=-(j-1)^2/2$, $x$ is fixed by $x_0=\sqrt{j}$.
When a test particle approaches the de Sitter horizon, the elapsed coordinate 
time diverges as
\begin{eqnarray}
t-t_0 &=& \lim_{x\rightarrow 1^-} \frac{\sqrt{2}l}{1-j^2}
\sqrt{\frac{1-j^2}{1-x^2}-1} \nonumber\\
&\sim& \lim_{x\rightarrow 1^-} \frac{1}{\sqrt{1-x}}\rightarrow +\infty .
\end{eqnarray}

\begin{figure}[ht]
\centerline{\psfig{figure=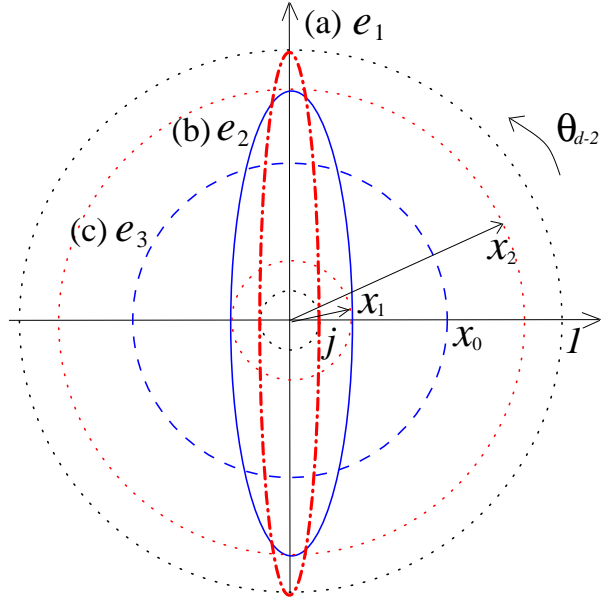,height=80mm}}
\caption{Orbits on ($x$, $\theta_{d-2}$)-plane :
(a) dash-dot line for $e_1$, (b) solid line for $e_2$,
 (c) dashed line for $e_3$. Circles stand for de Sitter horizon, aphelion,
 circular orbit, two perihelia from the outside.}
\label{orbit}
\end{figure}

By combining Eq.~(\ref{fixb4}) with Eq.~(\ref{geint}) and its integration,
we obtain an elliptic orbit equation for $e=e_2$ :
\begin{equation}\label{oesol}
\sin(2\theta_{d-2}+\theta_0)=\frac{-2j^2+(1+j^2+2e)x^2}
{x^2\sqrt{[(1+j)^2+2e][(1-j)^2+2e] }}.
\end{equation}
If we choose $x$ as the perihelion at $\theta_{d-2}=0$, $\theta_0$ becomes 
$-\pi/2$. From Eq.~(\ref{semi}), the semimajor axis $a$ of $x^2$ 
is given by
\begin{equation}
a=\frac{x_1^2+x_2^2}{2}=\frac{j^2+1+2e}{2}.
\end{equation} 
Eccentricity $\varepsilon$ of the ellipse  can be written 
\begin{equation}
\varepsilon=\sqrt{1-\frac{j^2}{a^2}},
\end{equation} 
Dependence of the orbit for $\varepsilon$ is the followings :
\begin{equation}
\left\{
\begin{array}{lll}
\varepsilon<1, &\;\;\;e \leq 0:\;\;\;& {\rm ellipse}, \\ 
\varepsilon=0, &\displaystyle \;\;\;e = -\frac{(j-1)^2}{2}:\;\;& {\rm circle}.
\end{array}
\right.
\end{equation}
This scheme agrees with qualitative discussion by using the effective 
potential (\ref{efpo}) and the energy diagram in Fig.~\ref{geo4}.
In terms of $A$ and $\varepsilon$,  Eq.~(\ref{oesol}) is rewritten by
\begin{equation}\label{gesan}
x=\sqrt{\frac{a(1-\varepsilon^2)}{1+\varepsilon\cos(2\theta_{d-2})}}. 
\end{equation} 
Eq.~(\ref{gesan}) follows $x=\sqrt{a(1-\varepsilon)}$ at $\theta_{d-2}=0$
 and $x=\sqrt{a(1+\varepsilon)}$ at $\theta_{d-2}=\pi/2$ as expected from
Eq.~(\ref{semi}). 
As shown in Fig.~\ref{orbit}, Eq.~(\ref{gesan}) satisfies the condition 
for closed orbits so-called Bertrand's theorem, which means a particle 
retraces its own foot step.

\setcounter{equation}{0}
\section{Penrose Diagram}\label{pendi}
Let us begin with a spacetime with physical metric $g_{\mu\nu}$, and
introduce another so-called unphysical metric $\bar{g}_{\mu\nu}$, 
which is conformally related to $g_{\mu\nu}$ such as
\begin{equation}
\bar{g}_{\mu\nu}=\Omega^{2}g_{\mu\nu}.
\end{equation}
Here, the conformal factor $\Omega$ is suitably chosen to bring in the 
points at infinity to a finite position so that the whole spacetime 
is shrunk into a finite region called Penrose diagram. 
A noteworthy property is that the null geodesics of two conformally related
metrics coincide, which determine the light cones and, in turn, define
causal structure.
If such process called conformal compactification is accomplished,
all the information on the causal structure of the de Sitter spacetime
is easily visualized through this Penrose diagram although distances
are highly distorted. In this section, we study detailed casual
structure in various coordinates in terms of the Penrose diagram.
Since every Penrose diagram is drawn as a two-dimensional square in the
flat plane, each point in the diagram denotes actually a $(d-2)$-dimensional
sphere $S^{d-2}$ except that on left or right side of the diagram.

\subsection{Static coordinates}
Let us introduce a coordinate transformation from the static
coordinates (\ref{pcs}) to Eddington-Finkelstein coordinates
$(x^+,~x^-,~\theta_{a})$ such as
\begin{equation}\label{stoe}
x^{\pm} \equiv t \pm \frac {l}{2}
\ln \frac {1+\frac{r}{l}}{1-\frac{r}{l}},
\end{equation}
where the range of $x^{\pm}$ is $(-\infty,\infty)$. 
As expected, $r/l=0$ results in a timelike curve for a static object 
at the origin, $x^{\pm}=t$.
Then the metric in Eq.~(\ref{pcs}) becomes
\begin{equation}\label{efc}
ds^2 = - {\rm sech}^2 \left(\frac{x^+-x^-}{2l}\right) dx^+ dx^-
+ l^{2}\tanh ^2 \left(\frac{x^+-x^-}{2l}\right) d\Omega _{d-2} ^2.
\end{equation}
Though the possible domain of real $r/l$ corresponds to the interior region
of the de Sitter horizon [0,1) due to the logarithm of Eq.~(\ref{stoe}),
the metric itself (\ref{efc}) remains to be real for the whole range of
$r/l$ since $({\rm sech} [(x^+-x^-)/2l], {\rm tanh} [(x^+-x^-)/2l])$ 
has (1,0) at 
$r/l=0$, (0,1) at $r/l=1$, and $(-1,\infty)$ at $r/l=\infty$, so it covers
the entire $d$-dimensional de Sitter spacetime as expected.

In order to arrive at Penrose diagram of our interest, these coordinates
are transformed into Kruskal coordinates ($U$, $V$) by
\begin{equation}\label{etok}
U \equiv -e^{x^-/l}  ~\mbox{and}~ V \equiv e^{-x^+/l},
\end{equation}
and then the metric takes the form
\begin{equation}\label{kc}
ds^2 =  \frac{l^2}{(1-UV)^2}\left[- 4dUdV
+ \left(  {1 + UV} \right) ^2 d\Omega _{d-2} ^2\right] .
\end{equation}
The value of $UV$ has $-1$ at the origin ($r/l=0$), $0$ at the horizon 
($r/l=1$), and $1$ at infinity ($r/l= \infty$) by a relation 
$r/l=(1+UV)/(1-UV)$. 
In addition, another relation $-U/V=e^{2t/l}$ tells us that the line of $U=0$ 
corresponds to past infinity ($t=-\infty$) and $V=0$ does to future 
infinity ($t=\infty$).
Therefore, the entire region of the de Sitter spacetime is drawn by
a Penrose diagram which is a square bounded by $|UV|=1$. At the horizon, 
$r=l$ so that $r/l=(1+UV)/(1-UV)$ implies $UV=0$.
\begin{figure}[ht]
\centerline{\psfig{figure=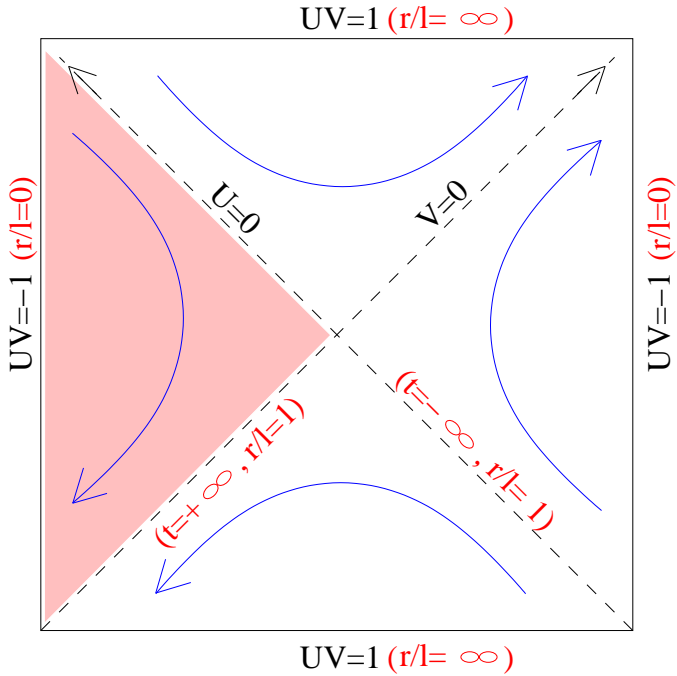,height=90mm}}
\caption{Penrose diagram of de Sitter spacetime from static coordinates.}
\label{pen1}
\end{figure}
Therefore, as shown in the Fig.~\ref{pen1}, 
the coordinate axes $U=0$ and $V=0$ (dashed lines) are nothing but the 
horizons and the arrows on those dashed lines stand for the directions of 
increasing $U$ and $V$.  The shaded region is the causally-connected 
region for the observer at the origin on the right-hand side of the square 
($UV=-1$, $r/l=0$). 

In the Kruskal coordinates (\ref{etok}), a Killing vector 
$\partial /\partial t$  in the static coordinates is expressed as
\begin{equation}
\frac{\partial}{\partial t} 
=\frac{U}{l}\frac{\partial}{\partial U}
-\frac{V}{l}\frac{\partial}{\partial V}.
\end{equation}
Thus norm of the Killing vector is
\begin{equation}
\left(\frac{\partial}{\partial t}\right)^2 = 4UV/(1-UV)^2. 
\end{equation}
Note that the norm of the Killing vector becomes null at $UV=0$.
In the region of $UV>0$, the norm is spacelike, while 
a half of the region with negative $V$ is forbidden. 
Timelike Killing vector is defined only in the shaded region with 
$\left(\partial/\partial t \right)^2 < 0 $
as shown in Fig.~\ref{pen1}.
Such existence of the Killing vector field $\partial/\partial t$
 guarantees conserved Hamiltonian which allows quantum mechanical description
 of time evolution. However, $\partial/\partial t$ is spacelike in
both top and bottom triangles and points toward the past in right triangle 
bounded by the southern pole in Fig.~\ref{pen1}.
Therefore, time evolution cannot be defined beyond the shaded region.
The absence of global definition of timelike Killing vector in the whole 
de Sitter spacetime may predict difficulties in quantum theory, 
e.g., unitarity.

\subsection{Conformal coordinates}
\begin{figure}[ht]
\centerline{\psfig{figure=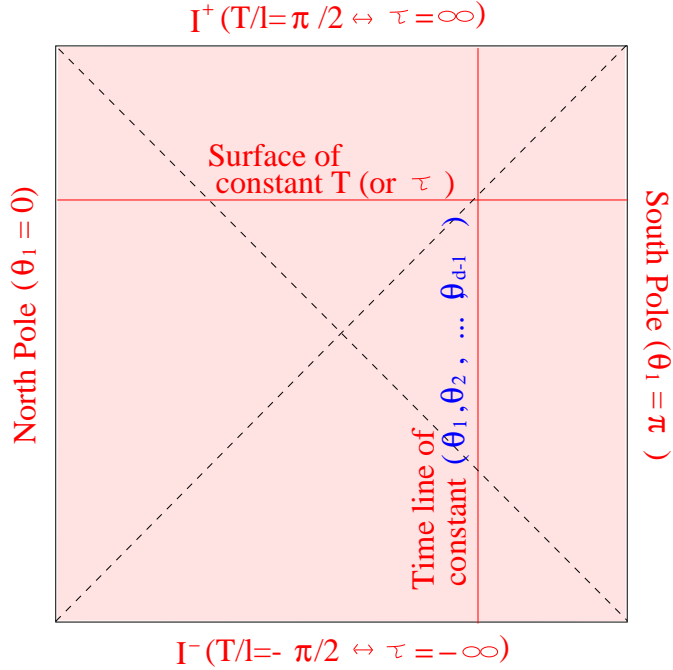,height=90mm}}
\caption{Penrose diagram of de Sitter spacetime in conformal 
coordinates.}
\label{pen3}
\end{figure}

As mentioned previously, the conformal metric (\ref{cct}) describes 
entire de Sitter spacetime and is flat except for a conformal factor 
$1/\cos^{2} (T/l)$ because of the scale symmetry (\ref{8}). These
properties are beneficial for drawing the Penrose diagram. By comparing
 the conformal metric (\ref{cct}) directly with that of the Kruskal coordinates
(\ref{kc}), we obtain a set of coordinate transformation
\begin{eqnarray}
U&=&\tan\left[\frac{1}{2}\left(\frac{T}{l}+\theta_{1}-\frac{\pi}{2}
\right)\right],\\
V&=&\tan\left[\frac{1}{2}\left(\frac{T}{l}-\theta_{1}+\frac{\pi}{2}
\right)\right],
\end{eqnarray}
where integration constants are chosen by considering easy comparison
with the quantities in the static coordinates. 
Since the range of conformal time $T$ is
$-\pi/2<T/l<\pi/2$, the horizontal slice $S^{d-1}$ at $T/l=-\pi/2
\Leftrightarrow \tau/l=-\infty$ ($T/l=\pi/2 \Leftrightarrow \tau/l=\infty$) 
forms a past (future) null infinity ${\cal I}^{-}$ 
(${\cal I}^{+}$) with $UV=1$ as shown in Fig.~\ref{pen3}. 
When $\theta_{1}=0$ $(\theta_{1}=\pi)$, it is a vertical line on 
the left (right) side with $UV=-1$, which is called by north (south) pole. 
When $T/l=-\theta_{1}+\pi/2$ $({\rm or}\; T/l=\theta_{1}-\pi/2)$, 
$U=0$ $({\rm or}\; V=0)$ 
corresponding to a null geodesic (or another null geodesic) starts 
at the south (or north) pole at the past null infinity ${\cal I}^-$ and 
ends at the 
north (or south) pole at the future null infinity ${\cal I}^+$, where all null
 geodesics originate and terminate (See two dashed lines at $45$ 
degree angles in Fig.~\ref{pen3}). Obviously, timelike surfaces are more 
vertical compared to the null geodesic lines and spacelike surfaces are 
more horizontal compared to those.
Therefore, every horizontal slice of a constant $T$ is a surface $S^{d-1}$
and every vertical line of constant $\theta'$s is timelike (See the horizontal
 and vertical lines in Fig.~\ref{pen3}).

\begin{figure}[ht]
\centerline{\psfig{figure=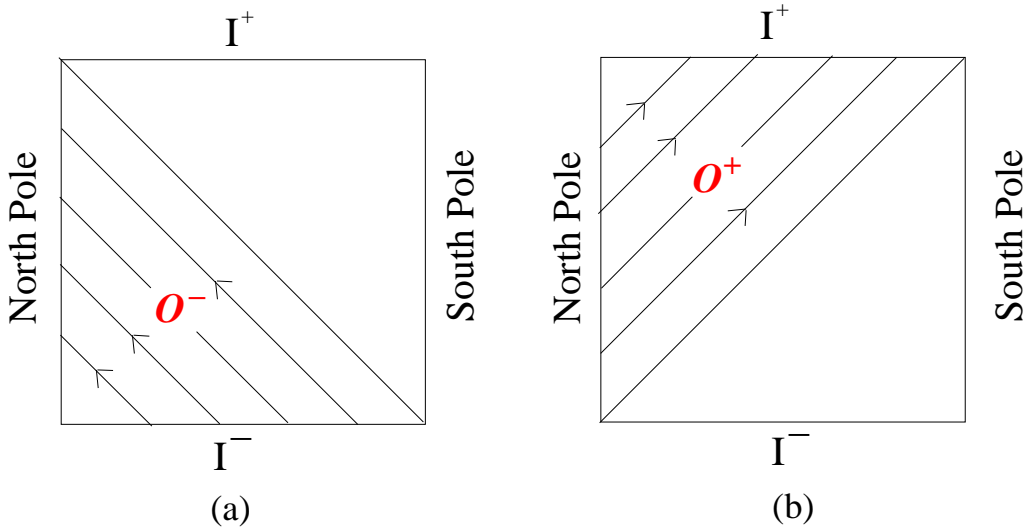,height=70mm}}
\caption{(a) Causal past of an observer at the north pole 
(b) Causal future of an observer at the north pole}
\label{pen4}
\end{figure}

Although this diagram contains the entire de Sitter spacetime, 
any observer cannot observe the whole spacetime. 
The de Sitter spacetime has particle horizon 
because past null infinity is spacelike, i.e.,
an observer at the north pole cannot see anything beyond his past
null cone from the south pole at any time as shown by the region ${\cal O}^-$
in Fig.~\ref{pen4}-(a), because the geodesics of particles are timelike.
The de Sitter spacetime has future event horizon because future null infinity 
is also spacelike~:  
The observer can never send a message to  any region beyond {${\cal O}^+$} 
as shown in Fig.~\ref{pen4}-(b). This fact is contrasted to the following from
 Minkowski spacetime where a timelike observer will eventually receive all 
history of the universe in the past light cone. Therefore, the fully accessible region 
to an observer at the north pole is the common region of both ${\cal O}^-$ 
and ${\cal O}^+$, which coincides exactly with the causally-connected region 
for the static observer at the origin in Fig.~\ref{pen1}. 

\subsection{Planar (inflationary) coordinates}
\begin{figure}[ht]
\centerline{\psfig{figure=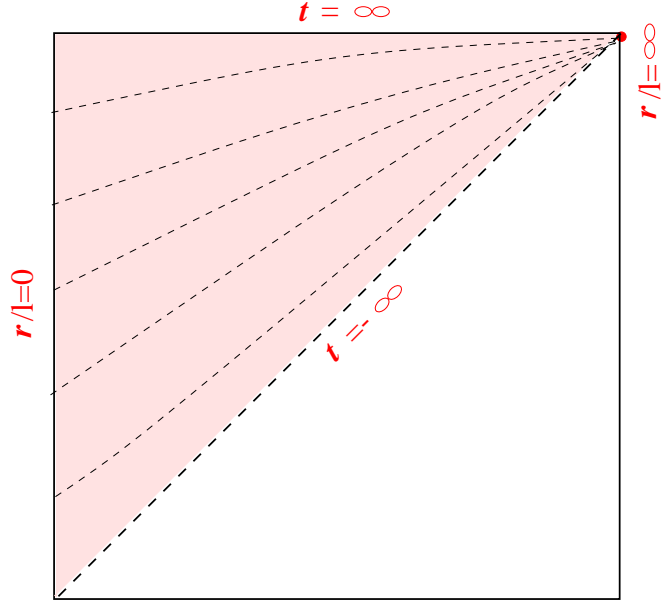,height=80mm}}
\caption{Penrose diagram of de Sitter spacetime in planar
coordinates.}
\label{pen2}
\end{figure}

By comparing the planar coordinates (\ref{plm}) with the Kruskal coordinates 
 (\ref{kc}), we again find a set of coordinate transformation 
\begin{eqnarray}
U &=& \frac{{\sf r}/l-e^{{- \sf t}/l}}{2}, \\ 
V &=& \frac{2}{e^{{- \sf t}/l} +{\sf r}/l}.\label{vv} 
\end{eqnarray}
Here we easily observe $V >0$ from Eq.~(\ref{vv}). 
$UV$ has $-1$ at both the origin ${\sf r}/l=0$ and infinity 
${\sf r}/l=\infty$ by a 
relation ${\sf r}/l= U+ 1/V$. In Fig.~\ref{pen2}, the origin 
corresponds to the boundary line at left side but 
the infinity to the point at upper-right corner.
According to another relation ${\sf t}/l=-\ln( 1/V-U)$,
past infinity ${\sf t}=-\infty$ has $V=0$ so it corresponds to the diagonal
 line, and future infinity 
${\sf t}=+\infty$ has $UV= 1$ so it corresponds to the horizontal line
 at upper side of the Penrose diagram. 
Therefore, the planar coordinates cover only the half of the de Sitter 
spacetime (See the shaded region in Fig.~\ref{pen2}).
 Every dashed line in Fig.~{\ref{pen2}} is a constant-time slice, 
which intersects with a line of ${\sf r}/l=0$ and is ($d-1$)=dimensional 
surface of infinite area with the flat metric of $k=0$. 
Although the whole de Sitter spacetime is geodesically complete as we discussed
 in the subsection~\ref{glogeo}, half of the de Sitter spacetime described by
the planar coordinates is incomplete in the past as shown manifestly in 
Fig.~\ref{pen2}. 

\setcounter{equation}{0}
\section{Discussion}

In this review, we have discussed classical geometry of 
$d$-dimensional de Sitter spacetime in detail~: Sections include
introduction of four representative coordinates (global (or closed),
conformal, planar (or inflationary), static), identification of
Killing vectors, geodesic motions, cosmological implication, 
and Penrose diagrams. 

Two important subjects are missing~: One is group theoretic approach
of the de Sitter spacetime and the other is energy because we do believe
they have to be systematically dealt with quantum theoretic topics 
like supersymmetry and Hamiltonian formulation of quantum theory. 
Once we try to 
supersymmetrize de Sitter group, inconsistency is easily encountered,
that either a supergroup including both unitary representations and 
isometries of the de Sitter spacetime is absent or any action compatible 
with de Sitter local supersymmetry contains vector ghosts~\cite{SUSY}.
In de Sitter spacetime two definitions are known to define the mass
of it~: One is the Abbott-Deser mass given as the eigenvalue of
zeroth component of Virasoro generators
and the other is the quasilocal mass obtained from Brown-York stress
tensor associated to a boundary of a spacetime~\cite{ener}.
However, unitarity and conserved energy are not satisfied 
simultaneously in quantum theory in the de Sitter spacetime,
and structure of de Sitter vacuum seems to ask more 
discussion~\cite{BD,covac,KC}.

Strikingly enough recent cosmological observations suggest the existence 
of extremely-small positive cosmological constant, 
however it just leads to a new form of
terrible cosmological constant problem. In the present stage, it seems to
force us to find its solution by understanding
quantum gravity (in de Sitter spacetime). Obviously the topics we reviewed form
a basis for the following research subjects still in progress, e.g., de Sitter
entropy and thermodynamics, dS/CFT correspondence, trans-Planckian 
cosmology, understanding of vacuum energy in the context of string theory, 
and etc, of which ultimate goal should be perfect understanding
of the cosmological constant problem as consequence of constructed quantum
gravity.

\appendix

\setcounter{equation}{0}
\section{Various Quantities}

For convenience,
various formulas and quantities which have been used in the previous
sections are summarized in the following four subsections.

\subsection{Global (closed) coordinates}

From the metric (\ref{gcm}) of the global coordinates ($\tau, \theta_i$),
 nonvanishing components of covariant metric $g_{\mu\nu}$ are
\begin{equation}\label{yd1}
g_{\tau\tau}=-1 ~~\mbox{and}~~ g_{ii}=l^2f^2 
\prod_{j=1}^{i-1}\sin^2\theta_j \;,
\end{equation}
and its inverse $g^{\mu\nu}$ has
\begin{equation}\label{yd2}
g^{\tau\tau}=-1 ~~\mbox{and}~~ g^{ii}=1/\left(l^2f^2 
\prod_{j=1}^{i-1}\sin^2\theta_j\right)\;.
\end{equation}
Jacobian factor $g$ of the metric (\ref{yd1}) becomes
\begin{equation}
g =-(l^2f^2)^{d-1}\prod_{j=1}^{d-1}
\prod_{i=1}^{j-1}\sin^2\theta_i \;.
\end{equation}
Nonvanishing components of the connection $\Gamma ^\mu _{\nu\rho}$ 
are
\begin{eqnarray}\label{chsy} 
& \displaystyle 
\Gamma^{i}_{\tau i} = \frac{1}{f}\left(\frac{df}{d\tau}\right) \;,
~~~\Gamma^{\tau}_{ii} = l^2f\left(\frac{df}{d\tau}\right)
\prod_{j=1}^{i-1}\sin^2\theta_j \;, & \nonumber\\
& \displaystyle 
\Gamma^{i}_{jj} = -\sin\theta_i\cos\theta_i\prod_{j=i+1}^{j-1}
\sin^2\theta_j \;,
~~~\Gamma^{i}_{ij} = \frac{\cos\theta_j}{\sin\theta_j} \;.&
\end{eqnarray}
For completeness, we mention non-zero components of the Riemann 
curvature tensor $R^{\mu}_{\;\nu\rho\sigma}$, 
which are consistent with Eq.~(\ref{9}) :
\begin{eqnarray}\label{riet}
& \displaystyle 
R^{\tau}_{\; i \tau i} = l^2f\left(\frac{d^2f}{d\tau^2}\right)
\prod_{j=1}^{i-1}\sin^2\theta_j \;,  
~~~R^{i}_{\; \tau i \tau} = -\frac{1}{f}\left(\frac{d^2f}{d\tau^2}\right) 
\;,& \nonumber\\
& \displaystyle
R^{i}_{\; j i j} = \left[1+l^2\left(\frac{df}{d\tau}\right)^2\right]
\prod_{j=1}^{i-1}\sin^2\theta_j \;,&
\end{eqnarray}
and the Ricci tensor 
$R_{\mu\nu}$ : 
\begin{eqnarray}\label{rite}
R_{\tau \tau} &=& -(d-1)\frac{1}{f}\left(\frac{d^2f}{d\tau^2}\right) \;, 
\nonumber\\
R_{ii} &=& 
\left\{ l^2f\left(\frac{d^2f}{d\tau^2}\right)
+(d-2)\left[1+l^2\left(\frac{df}{d\tau}\right)^2\right]\right\} 
\prod_{j=1}^{i-1}\sin^2\theta_j
\;.
\end{eqnarray}
Curvature scalar $R$ coincides with the formula in Eq.~(\ref{csgo}).

\subsection{Conformal coordinates}

From the metric (\ref{cmet}) of the conformal coordinates ($T, \theta_i$),
 nonvanishing components of covariant metric $g_{\mu\nu}$ are
\begin{equation}\label{zd1}
g_{T T}=-F^2 ~~\mbox{and}~~ g_{ii}=l^2F^2 
\prod_{j=1}^{i-1}\sin^2\theta_j \;,
\end{equation}
and its inverse $g^{\mu\nu}$ has
\begin{equation}\label{zd2}
g^{TT}=-1/F^2 ~~\mbox{and}~~ g^{ii}=1/\left(l^2F^2 
\prod_{j=1}^{i-1}\sin^2\theta_j\right)\;.
\end{equation}
Jacobian factor $g$ of the metric (\ref{zd1}) becomes
\begin{equation}
g =-\frac{(lF)^{2d}}{l^2}\prod_{j=1}^{d-1}
\prod_{i=1}^{j-1}\sin^2\theta_i \;.
\end{equation}
Nonvanishing components of the connection $\Gamma^{\mu}_{\nu\rho}$ are
\begin{eqnarray}\label{csyc} 
& \displaystyle \Gamma^{T}_{TT} = \Gamma^{i}_{T i} = 
\frac{1}{F}\left(\frac{dF}{dT}\right)\;,  
~~~\Gamma^{T}_{ii} = \frac{l^2}{F}\left(\frac{dF}{dT}\right)
\prod_{j=1}^{i-1}\sin^2\theta_j \;,& \nonumber\\
& \displaystyle 
\Gamma^{i}_{jj} = -\sin\theta_i\cos\theta_i\prod_{j=i+1}^{j-1}
\sin^2\theta_j \;,
~~~\Gamma^{i}_{ij} = \frac{\cos\theta_j}{\sin\theta_j} \;.&
\end{eqnarray}
Non-zero components of the Riemann curvature tensor 
$R^{\mu}_{\;\nu\rho\sigma}$ consistent with Eq.~(\ref{9}) are
\begin{eqnarray}\label{rtco}
R^{T}_{\; i T i} &=& \frac{l^2}{F^2}\left(F\frac{d^2F}{dT^2}-\frac{dF}{dT}
\right)\prod_{j=1}^{i-1}\sin^2\theta_j \;, \nonumber\\ 
R^{i}_{\; T i T} &=& -\frac{1}{F^2}\left[F\frac{d^2F}{dT^2}-
\left(\frac{dF}{dT}\right)^2 \right] \;,\nonumber\\
R^{i}_{\; j i j} &=& \frac{1}{F^2}\left[F^2+\left(\frac{dF}{dT}\right)^2
\right] \prod_{j=1}^{i-1}\sin^2\theta_j \;,
\end{eqnarray}
and non-zero components of the Ricci tensor $R_{\mu\nu}$ are 
\begin{eqnarray}\label{ritec1}
R_{TT} &=& -(d-1)\frac{1}{F^2}\left[ F\frac{d^2F}{dT^2}-
\left(\frac{dF}{dT}\right)^2\right] \;, \nonumber\\
R_{ii} &=& \frac{1}{l^2F^2} \left[ F\left(\frac{d^2F}{dT^2}\right)
+(d-3)\left(\frac{dF}{dT}\right)^2+(d-2)F^2 \right] 
\prod_{j=1}^{i-1}\sin^2\theta_j \;.
\end{eqnarray}
Therefore, curvature scalar $R$ in the conformal coordinates coincides 
with the formula in Eq.~(\ref{csco}).

\subsection{Planar (inflationary) coordinates}

From the metric (\ref{plme}) of the planar coordinates (${\sf t}, x_i$),
 nonvanishing components of covariant metric $g_{\mu\nu}$ are
\begin{equation}\label{ad1}
g_{\sf t t}=-1 ~~\mbox{and}~~ g_{ij}=a(\sf t)^2\gamma_{ij} \;, 
\end{equation}
and its inverse $g^{\mu\nu}$ has
\begin{equation}\label{ad2}
g^{\sf t t }=-1 ~~\mbox{and}~~ g^{ij}=\gamma^{ij}a^{-2} \;,
\end{equation}
where $\gamma^{ij}\gamma_{jk}= \delta^i_k$.
Jacobian factor $g$ of the metric (\ref{ad1}) becomes
\begin{equation}
g =-a^{2(d-1)}\gamma \;,
\end{equation}
where $\gamma=det \gamma_{ij}$.
Nonvanishing components of the connection $\Gamma^{\mu}_{\nu\rho}$ are
\begin{equation}\label{csyp} 
 \Gamma^{\sf t}_{ij} = a\left(\frac{da}{d{\sf t}}\right)\gamma_{ij}\;, 
~~~\Gamma^{i}_{{\sf t}j} = \frac{1}{a}\left(\frac{da}{d{\sf t}}\right)
\delta^i_j \;,
~~~\Gamma^{i}_{jk}=\frac{\gamma^{il}}{2}(\partial_j\gamma_{lk}+\partial_k 
\gamma_{lj}-\partial_l\gamma_{jk}) \;. 
\end{equation}
Non-zero components of the Riemann curvature tensor 
$R^{\mu}_{\;\nu\rho\sigma}$ consistent with Eq.~(\ref{9}) are
\begin{eqnarray} \label{rtpl}
& \displaystyle { R^{\sf t}_{\; i {\sf t} i} = a\left(\frac{d^2a}{d{\sf t}^2} \right)\;,
~~~ R^{i}_{\; {\sf t} i {\sf t}}  = -\frac{1}{a} 
\left( \frac{d^2a}{d{\sf t}^2} \right) \;,} & \nonumber\\ 
& \displaystyle{ R^{i}_{\; j k l} =\left[ \left(\frac{da}{d{\sf t}}\right)^2+k\right]
\left(\delta^i_k\gamma_{jl}-\delta^i_l\gamma_{kj} \right) \;,
~~~R^{\sf t}_{\; i j k} = a  \left(\frac{da}{d{\sf t}}\right)
\left[ \partial_j\gamma_{ki}-\partial_k\gamma_{ji}
+\partial_i\gamma_{jk}\right] \;,}&
\end{eqnarray}
and non-zero components of the Ricci tensor $R_{\mu\nu}$ are 
\begin{equation}\label{ritec2}
R_{\sf tt} = -\frac{d-1}{a}\left( \frac{d^2a}{d{\sf t}^2} \right) 
\;, ~~~R_{ij} =\left[ a\left(\frac{d^2a}{d{\sf t}^2}\right)+
(d-2)\left(\frac{da}{d\sf t}\right)^2+(d-2)k\right]\gamma_{ij}  \;. 
\end{equation}
Therefore, curvature scalar $R$ in the planar coordinates coincides with 
the formula in Eq.~(\ref{cspl}).

\subsection{Static coordinates}

From the metric (\ref{pc}) of the static coordinates ($t, r, \theta_a$),
 nonvanishing components of covariant metric $g_{\mu\nu}$ are
\begin{equation}\label{bd1}
g_{t t}=-A(r) e^{2\Omega(r)}\;,~~~~~  g_{rr}=\frac{1}{A(r)}\;,~~~~~
g_ {\theta_a \theta_a} = r^2 \prod_{a=1}^{b-1} \sin^2\theta_a \;,
\end{equation}
and its inverse $g^{\mu\nu}$ has
\begin{equation}\label{bd2}
g^{ t t }=-\frac{e^{-2\Omega}}{A}\;, ~~~~~ g^{ r r }=A\;, ~~~~~ 
g^{\theta_a \theta_a}=1/\left(r^2\prod_{a=1}^{b-1} \sin^2\theta_a 
\right)\;.
\end{equation}
Jacobian factor $g$ of the metric (\ref{bd1}) becomes
\begin{equation}
g =-e^{2\Omega}r^{2(d-2)}\prod_{b=1}^{d-2}
\prod_{a=1}^{b-1}\sin^2\theta_a \;.
\end{equation}
Nonvanishing components of the connection $\Gamma^{\mu}_{\nu\rho}$ are
\begin{eqnarray}\label{csys} 
&\displaystyle \Gamma^{t}_{rt} = \frac{1}{2A}\left(\frac{dA}{dr}
+2A\frac{d\Omega}{dr} \right)\;, 
~~~\Gamma^{r}_{tt} = \frac{Ae^{2\Omega}}{2}\left(\frac{dA}{dr}
+2A\frac{d\Omega}{dr} \right)\;,& \nonumber\\ 
& \displaystyle
\Gamma^{r}_{rr} = -\frac{1}{2A}\left(\frac{dA}{dr}\right)\;, 
~~~\Gamma^{\theta_a}_{r\theta_a} = \frac{1}{r}\;, 
~~~\Gamma^{\theta_a}_{\theta_b\theta_a} = 
\frac{\cos\theta_b}{\sin\theta_b}\;, & \nonumber\\ 
& \displaystyle
\Gamma^{r}_{\theta_a\theta_a} = -rA\prod_{a=1}^{a-1}\sin^2\theta_a \;, 
~~~\Gamma^{\theta_a}_{\theta_b\theta_b} = -\sin\theta_a\cos\theta_a
\prod_{a=a+1}^{b-1}\sin^2\theta_a \;.&
\end{eqnarray}
Non-zero components of the Riemann curvature tensor 
$R^{\mu}_{\;\nu\rho\sigma}$ consistent with Eq.~(\ref{9}) are
\begin{eqnarray}\label{rtst}
& \displaystyle 
R^{r}_{\; t r t} = \frac{Ae^{2\Omega}}{2} \left[ 3\left(\frac{dA}{dr}
\right)\left(\frac{d\Omega}{dr}\right)+2A\left(\frac{d\Omega}{dr}
\right)^2+\frac{d^2A}{dr^2}
+2A\left(\frac{d^2\Omega}{d^2r}\right) \right]\;,& \nonumber\\
& \displaystyle 
R^{t}_{\; r t r} = -\frac{1}{2A}\left[ 3\left(\frac{dA}{dr}
\right)\left(\frac{d\Omega}{dr}\right)+2A\left(\frac{d\Omega}{dr}
\right)^2+\frac{d^2A}{dr^2} +2A\left(\frac{d^2\Omega}{d^2r}\right) 
  \right]\;,& \nonumber\\
& \displaystyle
 R^{\theta_a}_{\; r \theta_a r} = -\frac{1}{2rA}\left(\frac{dA}{dr} 
\right) \;,  
~~~R^{t}_{\; \theta_a t \theta_a} = -\frac{r}{2}\left[ \frac{dA}{dr}
+2A\left(\frac{d\Omega}{dr}\right)\right]
\prod_{a=1}^{a-1}\sin^2\theta_a \;,& \nonumber\\
& \displaystyle 
R^{r}_{\; \theta_a r \theta_a} = -\frac{r}{2}
\left(\frac{dA}{dr} \right) \prod_{a=1}^{a-1}\sin^2\theta_a \;, 
~~~R^{\theta_b}_{\; \theta_a \theta_b \theta_a} =
(1-A) \prod_{a=1}^{a-1}\sin^2\theta_a \;,& \nonumber\\ 
& \displaystyle
R^{\theta_a}_{\; t \theta_a t} = \frac{Ae^{2\Omega}}{2r}
\left[\frac{dA}{dr}+2A\left(\frac{d\Omega}{dr}\right) \right]
\;.&  
\end{eqnarray}
Non-zero components of the Ricci tensor $R_{\mu\nu}$ are 
\begin{eqnarray}
& \displaystyle 
R_{tt} = \frac{Ae^{2\Omega}}{2r}\left\{(d-2)\left[\frac{dA}{dr}+
2A\left(\frac{d\Omega}{dr}\right) \right]+r\left[
3\left(\frac{dA}{dr} \right)\left(\frac{d\Omega}{dr}\right)
+2A\left(\frac{d\Omega}{dr} \right)^2+\frac{d^2A}{dr^2}
+2A\left(\frac{d^2\Omega}{d^2r}\right) \right]\right\} \;,& \nonumber\\
& \displaystyle
R_{rr} = -\frac{1}{2Ar} \left\{ (d-2) \left( \frac{dA}{dr} \right)
+ r \left[ 3\left( \frac{dA}{dr} \right) \left( \frac{d\Omega}{dr}
\right) + 2A \left( \frac{d\Omega}{dr} \right)^2+\frac{d^2A}{dr^2}
+2A\left(\frac{d^2\Omega}{d^2r}\right) \right] \right\} \;, & \nonumber
\end{eqnarray}
\begin{equation}
R_{\theta_a \theta_a} = \frac{r^2}{d-2}\left\{ \frac{d-2}{r^{d-2}}
\frac{d}{dr}\left[ r^{d-3}(1-A)\right]-A\frac{d-2}{r}\frac{d\Omega}{dr} 
\right\} \prod_{b=1}^{a-1}\sin^2\theta_b \;. 
\end{equation}
Therefore, curvature scalar $R$ in the static coordinates coincides with 
the formula in Eq.~(\ref{csst}). 
Here we obtain also simple derivative terms for the Einstein equations
\begin{equation}\label{sust}
\frac{R_{tt}}{A^2e^{2\Omega}}+R_{rr}=\frac{d-2}{r}\frac{d\Omega}{dr} \;,
\end{equation}
\begin{equation} \label{swst}
\frac{R_{tt}}{Ae^{2\Omega}}+AR_{rr}+\frac{(d-2)R_{ \theta_a \theta_a}}
{r^2 \displaystyle{\prod_{b=1}^{a-1}\sin^2\theta_b}}=\frac{d-2}{r^{d-2}}\frac{d}{dr}
\left[r^{d-3}(1-A)\right] \;.
\end{equation}

\section*{Acknowledgments}
Y.~Kim would like to thank M. Spradlin for helpful discussions.
This work was the result of research activities (Astrophysical Research 
Center for the Structure and Evolution of the Cosmos (ARCSEC)              
and the Basic Research Program, R01-2000-000-00021-0)
supported by Korea Science $\&$ Engineering Foundation.

\end{document}